\documentclass[12pt]{article}

\usepackage[superscript]{scicite}
\usepackage{graphicx}
\usepackage{times}

\usepackage{makecell}
\usepackage{diagbox}

\usepackage{xcolor}
\usepackage{url}

\usepackage{xr}
\makeatletter
\usepackage{slashbox}
\usepackage{indentfirst} 
\newcommand*{\addFileDependency}[1]{
  \typeout{(#1)}
  \@addtofilelist{#1}
  \IfFileExists{#1}{}{\typeout{No file #1.}}
}
\makeatother

\newcommand*{\myexternaldocument}[1]{
    \externaldocument{#1}
    \addFileDependency{#1.tex}
    \addFileDependency{#1.aux}
}
\myexternaldocument{supp}

\newcommand{\zcite}[1]{\scalebox{1.5}[1.5]{\raisebox{-0.84ex}{\cite{#1}}}} %%yiluo

\newenvironment{sciabstract}{
\begin{quote} \bf}
{\end{quote}}

\title{Enhancing charge stability of Ge quantum well heterostructures via SiGe layer composition engineering}

\author
{Ding-Ming Huang$^{1,\ast}$, Jun-Hang Liu$^{1,2,3 \ast}$, Han Gao$^{1}$, Jie-Yin Zhang$^{4}$,\\
Jian-Huan Wang$^{1}$, Fang-Ze Liu$^{2}$, Xin-Yu Zhou$^{2}$, Yi Luo$^{1}$, Bin-Xiao Fu$^{2}$,\\ 
Xiao-Fei Liu$^{1}$, Ji-Yin Wang$^{1, \dag}$, Jian-Jun Zhang$^{2, \dag}$, H. Q. Xu$^{1,5 \dag}$\\
\\
\normalsize{$^{1}$Beijing Academy of Quantum Information Sciences, Beijing 100193, China}\\
\normalsize{$^{2}$Beijing National Laboratory for Condensed Matter Physics, Institute of Physics,} \\
\normalsize{Chinese Academy of Sciences, Beijing, 100190 China}\\
\normalsize{$^{3}$University of Chinese Academy of Sciences, Beijing 100049, China}\\
\normalsize{$^{4}$Dongguan Institute of Materials Science and Technology,}\\
\normalsize{Chinese Academy of Sciences, Dongguan, Guangdong 523808, China}\\
\normalsize{$^{5}$Beijing Key Laboratory of Quantum Devices, Peking University, Beijing 100871, China}\\
\normalsize{$^{\ast}$These authors contributed equally to this work.}\\ 
\normalsize{$^{\dag}$To whom correspondence should be addressed; E-mail:}\\ 
\normalsize{wang\_jy@baqis.ac.cn; jjzhang@iphy.ac.cn; hqxu@pku.edu.cn.}
}

\date{\today}

\begin{document}
\baselineskip=24pt
\maketitle

\begin{sciabstract}
Composition modulation is a powerful technique for designing materials with tailored properties, fueling the development of advanced semiconductor devices. In this work, we have implemented this technique into Ge quantum well heterostructures, offering a promising avenue to address the critical challenge of charge stability in spin qubit devices. Harnessing the atomic-scale precision of molecular beam epitaxy, we have engineered the band structure of the SiGe top barrier via graded composition modulation, thereby reducing charge accumulation states at the SiGe-dielectric interface and strengthening the effective confinement to the hole gases in the Ge quantum wells. The enhanced charge stability of composition-modulated SiGe/Ge quantum well heterostructures is confirmed in Hall devices, featuring an enlarged stable gate voltage range. We have further fabricated quantum dot devices from the composition-modulated SiGe/Ge quantum well heterostructures and observed remarkably low charge noise with an averaged amplitude of $0.46\,\mathrm{\mu eV}/\mathrm{\sqrt{Hz}}$ at $1\,\mathrm{Hz}$---the lowest reported value for Ge quantum wells grown on silicon. This exceptional charge stability of the quantum dots persists in the few-hole regime, with no observable voltage drift over $\sim$hours. With reduced charge noise and enhanced energy stability, composition-modulated SiGe/Ge heterostructures exhibit significant potential for applications in building high-performance quantum devices, including spin qubits with a long coherence time.

\end{sciabstract}

\section*{Introduction}
Spin qubits in silicon (Si) and germanium (Ge) quantum dots have emerged as a leading platform for building scalable quantum processors\cite{loss1998,vandersypen2017,scappucci2021,stano2022,burkard2023}. These group-IV quantum dots offer significant advantages of high natural abundance of zero-nuclear-spin isotopes\cite{Itoh2014} and compatibility with the state-of-the-art semiconductor manufacturing process\cite{zwerver2022qubits,neyens2024probing,steinacker2025}. In recent years, notable progress has been made in Si/Ge-based spin qubits, including one- \cite{philips2022universal,Takeda2022,weinstein2023,george202412,Park2025} and two-dimensional spin qubit arrays\cite{hendrickx2021four,wang2024operating,Unseld2025,Dijkema2026}, high-fidelity single-\cite{yoneda2018quantum,Lawrie2023,zhou2025high} and two-qubit gate operations\cite{noiri2022fast,xue2022quantum,Mills2022quantum,Tanttu2022quantum,wang2024operating}, and spin manipulations at elevated temperatures\cite{Petit2020,camenzind2022hole,huang2024high}. Within this landscape, hole spin qubits in Ge stand out due to their inherent insensitivity to nuclear spin fluctuations and fast electric spin manipulation via strong spin-orbit coupling\cite{Bulaev2007,kloeffel2011strong}. Accordingly, ultrafast spin manipulations have been reported in Ge nanowires\cite{Froning2021ultrafast,wang2022ultrafast,liu2023ultrafast} and advanced qubit processors have been achieved in Ge two-dimensional hole gases (2DHGs)\cite{hendrickx2021four,wang2024operating,Dijkema2026,jirovec2021singlet,Zhang2024fourST}. Most notably, a ten-qubit system with high-fidelity gate operations\cite{wang2024operating}, a 16 gate-defined quantum dot array\cite{borsoi2024shared}, and a 18-qubit modular array\cite{Dijkema2026} have been demonstrated in Ge 2DHGs. These advances highlight the compelling promise of Ge hole gases for building scalable spin qubit processors.

In Ge quantum dot devices, charge noise poses a great challenge to hole spin coherence due to the mediating role of spin-orbit coupling\cite{Bulaev2005}. Charge trapping in the SiGe-dielectric interface is considered to be the main source of charge noise in Ge quantum devices\cite{Hoex2008,su2017effects,meyer2023electrical,massai2024impact,MaarXiv2026}. Charge tunneling into or out of the interface induces fluctuations in the charge environment of nearby quantum dots. The situation is further exacerbated when electric gates are energized as the tunneling process becomes energetically more favorable, which can cause severe charge instability. Previous attempts to mitigate charge instability in SiGe/Ge quantum well (QW) heterostructures have focused on surface treatments\cite{sangwan2025impact,li2025impact} or capping layer engineering\cite{su2017effects}, but their effectiveness remains to be validated in advanced quantum devices. Operating hole spin qubits at selected magnetic field orientations offers an additional means of reducing the impact of charge noise\cite{Piot2022BAngle,Bassi2026BAngle}; however, the optimal field direction differs among individual hole qubits---rendering this strategy impractical for large-scale spin qubit systems. At present, charge noise remains the key bottleneck that restricts the operational performance of Ge hole spin qubits.\cite{hendrickx2024sweet,stehouwer2025exploiting}. Therefore, a fundamental innovation is urgently required to tackle the charge noise issue and unlock the hole spin coherence for large-scale integration.

In this work, we employ band structure engineering via composition modulation in SiGe/Ge QW heterostructures to enhance the charge stability of quantum devices. We use molecular beam epitaxy (MBE) to grow heterostructures with composition-graded SiGe top barriers, lowering the valence band edge progressively along the growth direction. The valence band remains below the Fermi level near the SiGe-dielectric interface even under considerable gate biases, which energetically prevents charge accumulation in this region and meanwhile enhances electrical confinement to the holes in the Ge QW. This combined effect suppresses the charge tunneling out of the QW and therefore improves the charge stability. Correspondingly, Hall devices fabricated from the composition-modulated (CM) heterostructures exhibit a nearly three-fold expansion in the stable gate voltage range relative to their counterparts without composition modulation. Quantum dot devices fabricated from the CM heterostructures are characterized, yielding an average noise amplitude of $\sqrt{S_{\epsilon}}$= 0.46$\,\mathrm{\mu eV}$/$\mathrm{\sqrt{Hz}}$ at $1\,\mathrm{Hz}$---the lowest reported value for Ge QWs grown on silicon substrates. In the few-hole regime, the charge state transition voltage in each quantum dot exhibits no observable drift over $\sim$hours, confirming the exceptional charge stability. Our results demonstrate that incorporating composition modulation into Si/Ge heterostructures represents a promising strategy for realizing low-charge-noise material stacks, thereby supporting advanced quantum devices with long coherence times.

\section*{Results}
\begin{figure}[!t] 
\centering
\includegraphics[width=\linewidth] {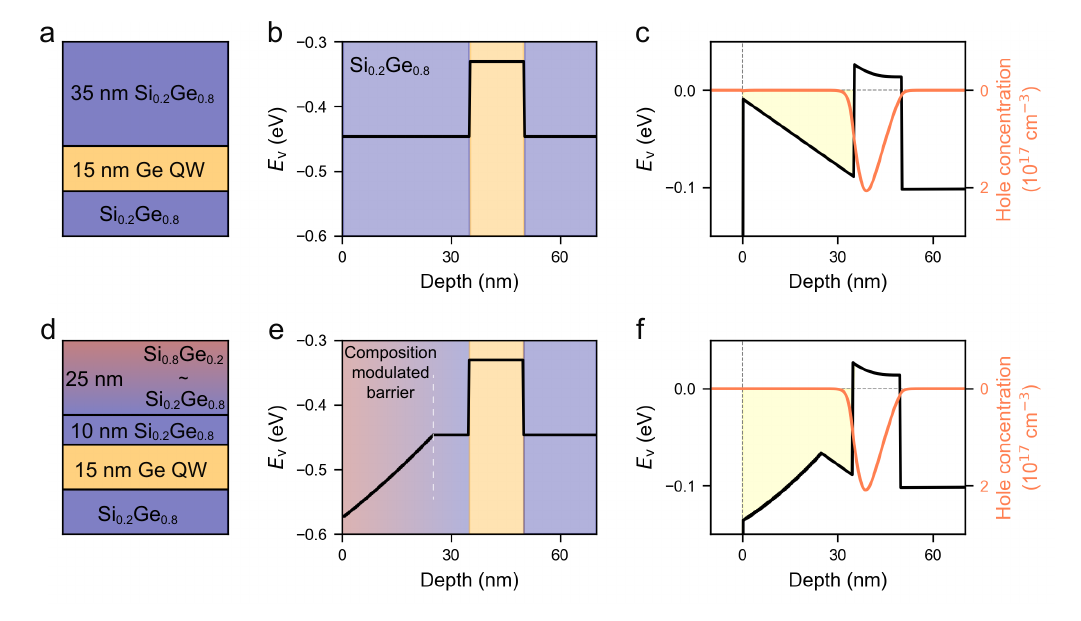}
\caption{\textbf{Layer structures, simulated energy band profiles and hole density distributions of SiGe/Ge QW heterostructures with a Si$_{0.2}$Ge$_{0.8}$ top barrier and a CM top barrier.} \textbf{a} and \textbf{d}, Layer structures of the heterostructures with a Si$_{0.2}$Ge$_{0.8}$ top barrier and with a CM top barrier layer. \textbf{b} and \textbf{e}, VBM energy ($E_\mathrm{v}$) profiles at zero electric field for the heterostructures in figures a and d, respectively. The white dashed line in figure e marks the interface between the Si$_{0.2}$Ge$_{0.8}$ spacer layer and the CM layer. \textbf{c}, $E_\mathrm{v}$ profile and hole density distribution of the heterostructures with a Si$_{0.2}$Ge$_{0.8}$ top barrier, i.e., figure a, at a gate voltage approaching to the onset of charge trapping. A triangular well forms at the SiGe-dielectric interface. \textbf{f}, $E_\mathrm{v}$ profile and hole density distribution of the CM SiGe/Ge QW heterostructures, i.e., figure d, under the same gate voltage as applied in figure c. The valence band of the top barrier is persistently away from the Fermi level and a substantial barrier remains for the holes in the QW.
}\label{fig1}
\end{figure}

In strained Ge QWs, SiGe barriers of constant Si and Ge compositions are generally employed to achieve effective potential confinement to the Ge hole gases, as shown in Fig. \ref{fig1}a. Figure \ref{fig1}b displays corresponding valence band diagram without external gate voltages with the Fermi level defined as the middle of the band gap of the Ge layer. Figure \ref{fig1}c presents the simulated energy band diagram and charge distribution at a finite gate voltage, which are obtained by solving self-consistent 1D Schrödinger-Poisson equation\cite{virgilio2006type,sant2013band,hebal2021general} (see details in Methods). At this gate voltage, the valence band maximum (VBM) at the SiGe–dielectric interface lies only 10 meV below the Fermi level. Then, charge accumulation at the interface becomes energetically available upon further negative biasing of the gate. Meanwhile, the applied gate voltage also leads to a tilted potential profile in the top SiGe barrier, which significantly reduces the confinement strength and enables holes to tunnel out from the Ge QWs. The combined effects ease charge movement across the SiGe barrier, leading to charge instability in the SiGe/Ge QW heterostructures. From the energy band perspective, lowering the valence band in the top barrier can be an effective approach to addressing the issues collectively. Figure \ref{fig1}d shows the layer structures of a Ge QW with a CM SiGe top barrier. On top of the Ge layer lies a SiGe spacing layer with constant Si and Ge compositions, above which is a CM SiGe layer with gradually increased Si composition. Then, the VBM progressively lowers in the CM SiGe layer as shown in Fig. \ref{fig1}e. As a result, the CM heterostructure exhibits significant resilience against the formation of charge accumulation at SiGe-dielectric interface. In Fig. \ref{fig1}f, the VBM of the top SiGe layer persists away from the Fermi level even with the gate energized (the same gate voltage is applied as in Fig. \ref{fig1}c). In this case, charge accumulation is energetically forbidden in the SiGe-dielectric interface. Concurrently, the lowered valence band enhances the charge confinement to the holes in the well. Therefore, a significant improvement in the charge stability of the CM SiGe/Ge QW heterostructure is expected.  

\begin{figure}[!t] 
\centering
\includegraphics[width=0.5\linewidth] {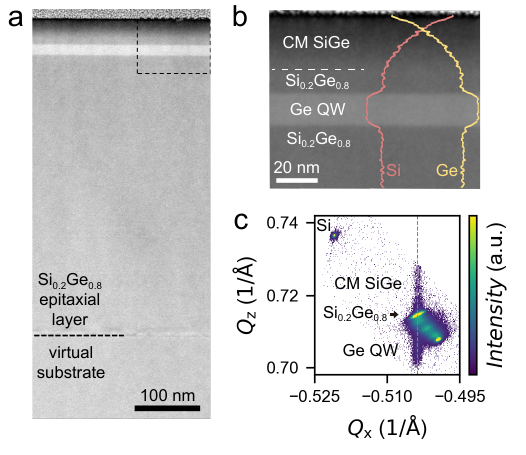}
\caption{\textbf{Structure characterizations on heterostructure H1.} \textbf{a}, Cross-sectional TEM image of the epitaxial layers in H1. \textbf{b}, HAADF TEM image of the QW region in the vicinity of the dashed rectangle marked in figure a. The overlaid traces show the composition profiles of Si and Ge along the growth direction measured by EDS. \textbf{c}, XRD-RSM image of H1. The vertical dashed line marks the uniform in-plane lattice constant of the epitaxial layers, demonstrating a fully strained condition in both the Ge QW and the CM top barrier layer.
}\label{fig2}
\end{figure}

Following the material stack design in Fig. \ref{fig1}, we have employed MBE to grow SiGe/Ge QW heterostructures with a Si$_\mathrm{0.2}$Ge$_\mathrm{0.8}$ top barrier (denoted as H0) and with a CM SiGe barrier (denoted as H1). See details of the material growth in Methods. As demonstrated in our previous work, MBE allows for the growth of high-quality SiGe/Ge QW heterostructures with atomically sharp interface and precise composition control\cite{zhang2024high}. The heterostructures are grown on a Si$_\mathrm{0.2}$Ge$_\mathrm{0.8}$ virtual substrate, which is epitaxially deposited on a Si(001) wafer via a reverse-graded-buffer technique (Supplementary Section I). Figure \ref{fig2} presents the material characterizations of the heterostructure H1, in which the top barrier comprises $10\,\mathrm{nm}$ Si$_\mathrm{0.2}$Ge$_\mathrm{0.8}$ and $25\,\mathrm{nm}$ graded Si$_\mathrm{x}$Ge$_\mathrm{1-x}$ layer (x varies from 0.2 to 0.8). Figure \ref{fig2}a shows a cross-sectional transmission electron microscopy (TEM) image of the epitaxial layers of H1. There are sharp interfaces in both the substrate and the QW regions, without observable threading dislocations. Figure \ref{fig2}b shows the magnified high-angle-annular-dark-field (HAADF) TEM image of the structures in the Ge QW region, taken from the vicinity of the dashed rectangle in Fig. \ref{fig2}a. Regions with higher Ge composition exhibit brighter contrast and the graded contrast at the top arises from the modulated compositions of Si and Ge. The overlaid traces (yellow for Ge and pink for Si) show the composition variation along the growth direction, as measured by energy-dispersive spectroscopy (EDS). Based on a comparative analysis of EDS and secondary ion mass spectrometry (SIMS) data (Supplementary Section II), the compositions in the CM layer varies from Si$_\mathrm{0.2}$Ge$_\mathrm{0.8}$ to Si$_\mathrm{0.8}$Ge$_\mathrm{0.2}$ as designed. Figure \ref{fig2}c shows an X-ray diffraction reciprocal space mapping (XRD-RSM) of H1. In the figure, two main intensity peaks arise from the (\(\bar{2}\bar{2}4\)) facets in the Si wafer and the Si$_\mathrm{0.2}$Ge$_\mathrm{0.8}$ layers, respectively. The signal from the Ge QW layer appears below the Si$_\mathrm{0.2}$Ge$_\mathrm{0.8}$ peak, while the signal from the CM SiGe layer is above it. This indicates that the Ge QW layer and the CM layer share the same in-plane lattice constants as the Si$_\mathrm{0.2}$Ge$_\mathrm{0.8}$ layers, and both are grown under in-plane strain. The reference heterostructure H0, with $35\,\mathrm{nm}$ Si$_\mathrm{0.2}$Ge$_\mathrm{0.8}$ as the top barrier, is characterized in the same way and the results confirm that H0 is also grown as designed (Supplementary Section III).

\begin{figure}[!t] 
\centering
\includegraphics[width=0.5\linewidth] {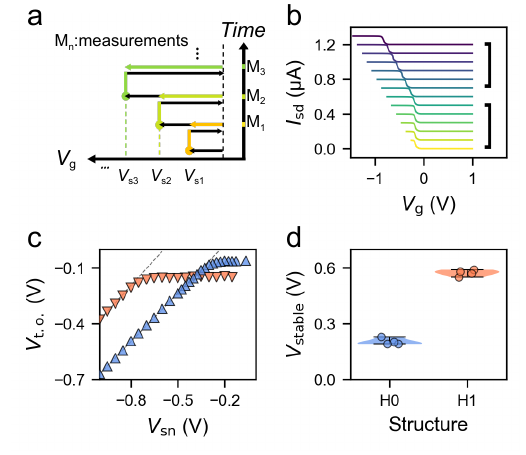}
\caption{\textbf{Stability characterization measurements of Hall devices.} \textbf{a}, Schematic diagram of measurement protocol on Hall bar devices. The ending gate voltage $V_\mathrm{sn}$ decrease over successive measuring loops. \textbf{b}, Transfer characteristic curves at varied $V_\mathrm{sn}$, distinguished by colors, for a representative device from H1 (H1-Dev1). These curves are vertically offset by 0.1 µA for clarity. The threshold voltages possess a “stable zone” for small $|V_\mathrm{sn}|$ and begin to drift linearly with more negative $V_\mathrm{sn}$ (“linear shift zone"). The two zones are marked by black brackets. \textbf{c}, Threshold voltages as a function of $V_\mathrm{sn}$ for H1-Dev1 (orange) and H0-Dev1 (blue). The dashed lines show the extrapolation of the threshold voltages in “linear shift zone”. \textbf{d}, The voltage span of the “stable zone", $V_\mathrm{stable}$, of Hall devices made from heterostructures H0 and H1. 
}\label{fig3}
\end{figure}

To characterize the charge stability of the heterostructures (H0 and H1), we have measured their electrical properties with Hall bar devices. For each heterostructure, a few Hall bar devices are fabricated and measured in a cryostat with a temperature of near $2\,\mathrm{K}$ (Methods). Magnetotransport measurements demonstrate that the 2DHGs in H1 and H0 exhibit comparable saturation mobilities and analogous integer quantum Hall effect characteristics (Supplementary Fig. S4). Figure \ref{fig3}a illustrates the stress measurement protocol for the Hall devices, which is used to quantify their charge stability\cite{massai2024impact,meyer2023electrical}. For the $n$-th cycle, the gate voltage is first held at $V_\mathrm{sn}$ for a period of time, then set to the starting point, and finally scanned from the starting point to $V_\mathrm{sn}$ to have a transfer characteristic curve. Figure \ref{fig3}b shows the transfer characteristic curves at varied $V_\mathrm{sn}$ for a representative device made from heterostructure H1 (H1-Dev1). These curves can be categorized into two groups: (i) For small $|V_\mathrm{sn}|$, the threshold voltages remain constant despite variations in $V_\mathrm{sn}$, corresponding to “stable zone" in the figure; (ii) At more negative $V_\mathrm{sn}$, the threshold voltages exhibit a linear shift with increasing $V_\mathrm{sn}$, corresponding to “linear shift zone". These voltage shifts are considered as a consequence of charge trapping at the SiGe-dielectric interface. Figure \ref{fig3}c presents the threshold voltages of turn-on curves, $V_\mathrm{t.o.}$, as a function of $V_\mathrm{sn}$ for Hall devices H1-Dev1 (orange) and H0-Dev1 (blue). By extrapolating the linear shift data, the stable gate voltage spans, $V_\mathrm{stable}$, are extracted (Methods). The same processing has been applied to four devices from H1 and four devices from H0 (Supplementary Fig. S5), and corresponding results are summarized in Fig. \ref{fig3}d. It is evident that heterostructure H1 shows a 2.75-fold expansion in stable gate range compared to H0, demonstrating an improved electrostatic stability in the CM heterostructure. This expansion is further validated with energy-band simulation (Supplementary Fig. S6). Due to the lowered VBM in the SiGe top barrier of H1, a higher gate voltage can be applied before the emergence of charge accumulation at the SiGe-dielectric interface.

\begin{figure}[!t] 
\centering
\includegraphics[width=\linewidth] {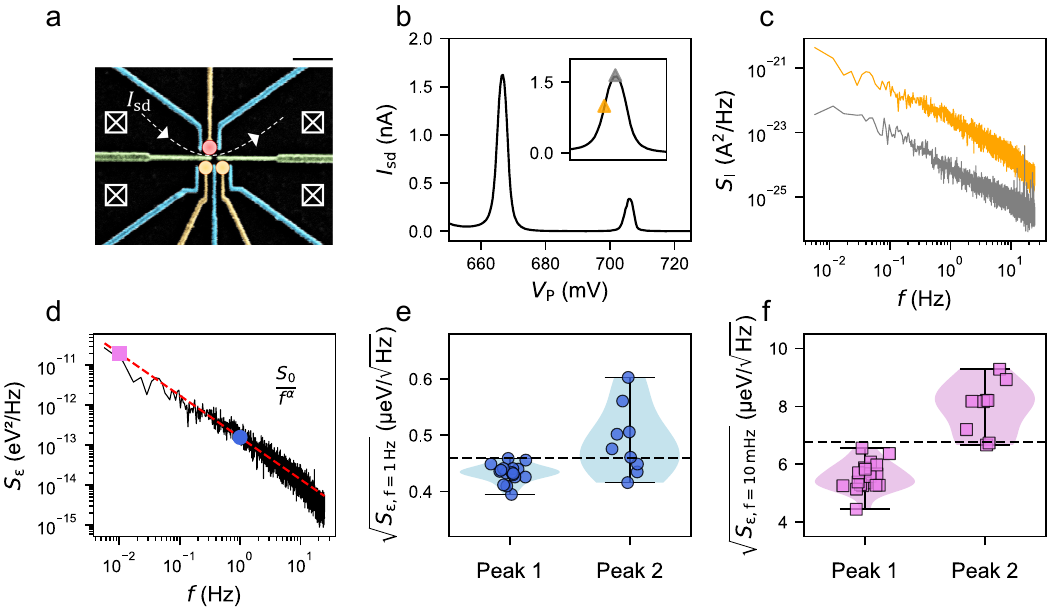}
\caption{\textbf{Charge noise measurements of quantum dot devices via current spectroscopy method.} \textbf{a}, False-color SEM image of device A, comprising a double quantum dot (yellow) and a charge sensor dot (red). The barrier gates (light blue) and the plunger gates (orange) are used to tune tunneling barriers and electro-chemical potential of quantum dots, respectively. The cutter gates (light green) are for separating the charge sensor dot from the double dot electrically. The scale bar is $300\,\mathrm{nm}$. \textbf{b}, Current through the charge sensor dot $I_\mathrm{sd}$ as a function of corresponding plunger gate voltage $V_\mathrm{P}$. Two coulomb oscillation peaks (peak 1 and peak 2) are used for charge noise measurements. Inset (zoom-in of peak 1): orange and grey triangles mark the flank and the apex of the Coulomb peak, respectively. \textbf{c}, Current spectral density $S_\mathrm{I}$ measured at the the flank and the apex of the Coulomb peak as marked in figure b. \textbf{d}, Noise power spectral density $S_\mathrm{\epsilon}$ converted from the orange trace in figure c and corresponding fitting curve (red dashed line) with the formula $S_\mathrm{0}/f^{\alpha}$. Here, $\alpha$ is fitted to be $1.05\,{\pm}\,0.01$. The blue dot and the violet square mark the fitted values of $S_\mathrm{\epsilon}$ at $1\,\mathrm{Hz}$ and $10\,\mathrm{mHz}$, respectively. \textbf{e}, Charge noise amplitudes at $1\,\mathrm{Hz}$ extracted from the measurements at the flanks of the two Coulomb peaks. \textbf{f}, Corresponding charge noise amplitudes at $10\,\mathrm{mHz}$. The distributions of violin plots in figures e and f are obtained from multiple current traces acquired at random time intervals, and the average noise amplitude of device A at $1\,\mathrm{Hz}$ and $10\,\mathrm{mHz}$ are 0.46 and 6.77$\,\mathrm{\mu eV}$/$\mathrm{\sqrt{Hz}}$, respectively, highlighted with black dashed lines.}\label{fig4}
\end{figure}

In Fig. \ref{fig4}, we further quantify the charge noise level of the CM Ge QW using quantum dot devices. Three quantum dot devices (device A-C) are made from H1 (see SEM images in Supplementary Fig. S7). Figure \ref{fig4}a shows a false-color SEM image of device A, comprising a charge sensor dot and a double quantum dot. Light blue, orange, and light green electrodes are barrier gates, plunger gates and cutter gates, respectively. The quantum dot devices are measured in a dilution refrigerator with a base temperature of about $60\,\mathrm{mK}$. In Figs. \ref{fig4}b-\ref{fig4}f, charge noise is measured using only the sensor dot. Figure \ref{fig4}b displays the Coulomb oscillation current $I_\mathrm{sd}$ of the sensor dot as a function of its plunger gate voltage $V_\mathrm{P}$. Then, the current noise of the dot is characterized at the apex and flank of the Coulomb peak as marked in the inset panel (see Methods). The corresponding current noise power spectral density $S_\mathrm{I}$ at the two positions is presented in Fig. \ref{fig4}c. The current noise at the flank of the peak (orange) is larger than that at the apex (grey) by nearly two orders of magnitude, due to its enhanced susceptibility to charge fluctuations. This indicates that the measured current noise at the flank of the Coulomb peak arises predominantly from the dot device. The charge noise spectral density $S_\mathrm{\epsilon}$ can be converted from $S_\mathrm{I}$ by taking into account the slope of current trace at the Coulomb peak flank and the lever arm of the plunger gate (see Methods). Following this procedure, $S_\mathrm{\epsilon}$ for Coulomb peak 1 is obtained and presented in Fig. \ref{fig4}d (see Supplementary Fig. S8 for the extraction of relevant parameters). The black trace presents $S_{\epsilon}$ as a function of frequency $f$ for Coulomb peak 1, calculated from the orange trace in Fig. \ref{fig4}c. The red dashed line is a fitting curve to the black trace with a formula $S_{0}/f^{\alpha}$ and $\alpha$ is extracted to be $1.05\,{\pm}\,0.01$. The nearly $1/f$ trend of the noise spectrum indicates that a group of two-level fluctuators exists near the sensor dot in the device. The charge noise amplitude $\sqrt{S_{\epsilon}}$ at $1\,\mathrm{Hz}$ and $10\,\mathrm{mHz}$ can be extracted as marked by the blue dot and the violet square in the figure. In this way, the charge noise amplitudes of the two Coulomb peaks at $1\,\mathrm{Hz}$ and $10\,\mathrm{mHz}$ are extracted and presented in Figs. \ref{fig4}e and \ref{fig4}f, respectively (see raw data in Supplementary Fig. S9). In the two figures, each cluster of discrete data points represents multiple measurement trials taken at some random time intervals. In general, peak 2 exhibits a broader distribution of noise amplitude at both $1\,\mathrm{Hz}$ and $10\,\mathrm{mHz}$, attributed to subtle variation of the charge environment with the plunger gate. As indicated by the black dashed lines, the average charge noise amplitude of the two peaks at $1\,\mathrm{Hz}$, $\sqrt{S_{\epsilon, f=1\,\mathrm{Hz}}}$, is about 0.46$\,\mathrm{\mu eV}$/$\mathrm{\sqrt{Hz}}$, and $\sqrt{S_{\epsilon, f=10\,\mathrm{mHz}}}$ is about 6.77$\,\mathrm{\mu eV}$/$\mathrm{\sqrt{Hz}}$. The noise amplitude of our device is lower than that reported previously in Ge QWs, except for a very recent work\cite{stehouwer2025exploiting}, which will be further discussed in the following.      

\begin{figure}[!t] 
\centering
\includegraphics[width=0.5\linewidth] {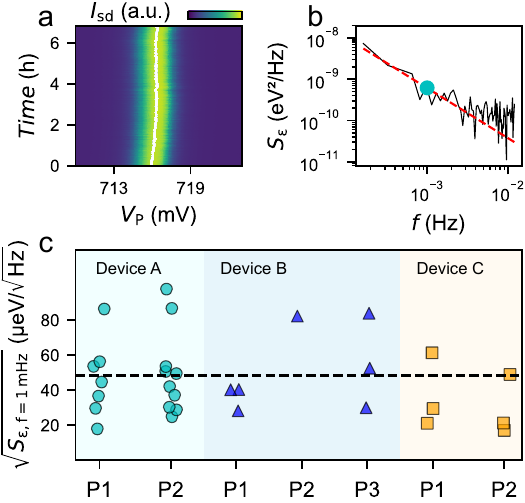}
\caption{\textbf{Charge noise measurements of quantum dot devices by Coulomb peak tracking method.} \textbf{a}, Repeated measurements of a Coulomb peak of the sensor dot in device A over 7 hours. The overlapped white line tracks the peak positions along $V_\mathrm{p}$ over time. \textbf{b}, Charge noise power spectral density extracted from figure a. The red dashed line shows the fitting to the data in the range of $0.2-2\,\mathrm{mHz}$ with the formula $S_{0}/f^{\alpha}$. The exponent value $\alpha$ is fitted to be $1.21\,{\pm}\,0.15$. The cyan dot marks the charge noise amplitude at $f=1\,\mathrm{mHz}$. \textbf{c}, Noise amplitude at $f=1\,\mathrm{mHz}$ for different Coulomb peaks of device A-C. The distributions of the results are from repeated measurements sampled at random time intervals. The average noise amplitude at $1\,\mathrm{mHz}$ is $46.48\,\mathrm{\mu eV/\sqrt{Hz}}$, highlighted with a black dashed line.}\label{fig5}
\end{figure}

\begin{figure}[!t] 
\centering
\includegraphics[width=0.5\linewidth] {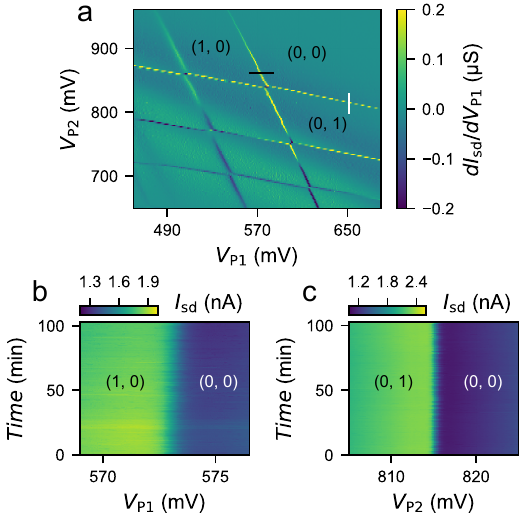}
\caption{\textbf{Charge stability diagram measurements of device A.} \textbf{a}, Charge stability diagram of the double quantum dot of device A in the few-hole regime. The numbers in the brackets represent the hole numbers in the two quantum dots. The black and white bars mark the gate voltage range scanned in figures b and c. \textbf{b} and \textbf{c}, Repeated measurements of sweeping $V_\mathrm{P1}$ / $V_\mathrm{P2}$ across the border of the charge state (0, 0) to (1, 0) / (0, 1).}\label{fig6}
\end{figure}

Coulomb peak tracking is another method generally used to quantify charge noise at low frequencies in quantum dot devices\cite{Kranz2020}. In Fig. \ref{fig5}, this method is employed to all three quantum dot devices (A-C) to obtain charge noise at frequency as low as $1\,\mathrm{mHz}$. Figure \ref{fig5}a is a representative 2D plot of Coulomb peak tracking results from the sensor dot of device A. Here, the Coulomb peak is repeated scanned over a time period of nearly 7 hours. The peak positions in $V_\mathrm{P}$ are extracted as a function of time and are seen as a white line overlapped on the 2D plot. Figure \ref{fig5}b presents the obtained charge noise power spectral density $S_{\epsilon}$ (Methods). 
The spectrum simply follows $1/f^{\alpha}$ in the frequency range of $0.2-2\,\mathrm{mHz}$ and shows saturation trend at higher frequencies.
The data in the range of $0.2-2\,\mathrm{mHz}$ are fitted with the form of $1/f^{\alpha}$ (red dashed line), where $\sqrt{S_{\epsilon}}$ at $1\,\mathrm{mHz}$ can be extracted as marked by the cyan dot. Meanwhile, the noise amplitude at higher frequencies can be obtained by extrapolating the fitting curve. In such a way, the charge noise amplitude of different Coulomb peaks of the three devices at $1\,\mathrm{mHz}$ is extracted and shown in Fig. \ref{fig5}c (see the raw data and the fittings in Supplementary Figs. S10 and S11). As indicated by the black dashed line, the average charge noise amplitude of all three devices at $1\,\mathrm{mHz}$ is $\sim$46.48$\,\mathrm{\mu eV}$/$\mathrm{\sqrt{Hz}}$.  

As a further demonstration of the charge stability of our devices over extended time periods, we have presented the charge stability diagram measurements of device A in Fig. \ref{fig6}. In Fig. \ref{fig6}a, the transconductance of charge sensor d$I_\mathrm{sd}$/d$V_\mathrm{P1}$ is measured as a function of the two plunger gate voltages (charge stability diagram). The numbers in brackets mark the hole numbers in the two quantum dots and different charge states are separated by clear boundaries. Repeated measurements of filling one hole into the left and right dots are performed by scanning corresponding plunger gates (black and white bars), and the results are presented in Figs. \ref{fig6}b and \ref{fig6}c. The gate voltages of the charge state transitions between (0, 0)-(1, 0) or (0, 0)-(0, 1) keep stable over 100 minutes, confirming the excellent charge stability of our Ge quantum dot devices.

\begin{table}
\begin{center}
\scalebox{0.75}
{
\setlength{\tabcolsep}{5pt}
\renewcommand{\arraystretch}{2.5}
\begin{tabular}{|l||*{8}{c|}}\hline
%\diagbox{}{} &
\backslashbox{Reference}{}&
\makecell{Material stack}&
\makecell{Substrate}&
\makecell{Barrier\\thickness}&
\makecell{Dielectric}&
\makecell{$\sqrt{S_{\epsilon, f=1\,\mathrm{Hz}}}$}&
\makecell{$\sqrt{S_{\epsilon, f=10\,\mathrm{mHz}}}$}\\\hline

{Hendrickx et al.\cite{hendrickx2018gate}}&
{Ge(18 nm)/Si$_{0.2}$Ge$_{0.8}$}&
{Si}&
{22 nm}&
\makecell{17 nm\\Al$_{2}$O$_{3}$}&
{1.4$^a$}&
{/}\\\hline

{Lodari et al.\cite{Lodari2021low}}&
{Ge(16 nm)/Si$_{0.2}$Ge$_{0.8}$}&
{Si}&
{55 nm}&
{Al$_{2}$O$_{3}$}&
{0.6$^a$}&
{/}\\\hline

{Stehouwer et al.\cite{stehouwer2025exploiting}}&
{Ge(16 nm)/Si$_{0.2}$Ge$_{0.8}$}&
{Ge}&
{55 nm}&
\makecell{14 nm\\Al$_{2}$O$_{3}$}&
{0.3$^a$}&
{5.5$^b$}\\\hline

{Massai et al.\cite{massai2024impact}}&
{Ge(20 nm)/Si$_{0.2}$Ge$_{0.8}$}&
{Si}&
{47 nm}&
\makecell{14 nm\\SiO$_{2}$}&
{/}&
{68.4$^b$}\\\hline

{Borovkov et al.\cite{Borovkov2026}}&
{Ge(18nm)Si$_{0.3}$Ge$_{0.7}$}&
{Si}&
{4 nm}&
\makecell{HfO$_{2}$ or \\Al$_{2}$O$_{3}$}&
{1.8$^a$}&
{/}\\\hline

{This work}&
{Ge(15 nm)/Si$_{0.2}$Ge$_{0.8}$/Si$_{x}$Ge$_{1-x}$}&
{Si}&
{35 nm}&
\makecell{10 nm\\Al$_{2}$O$_{3}$}&
{0.46$^a$}&
{6.77$^a$/5.92$^b$}\\\hline

\end{tabular}
}
\caption{\label{table:S2}\textbf{Reported charge noise levels in quantum dots devices made of different Ge QWs in recent, with $\sqrt{S_{\epsilon}}$ determined via the technique of current spectroscopy$^a$ or Coulomb peak tracking$^b$.} }

\end{center}
\end{table}

In order to make quantitative comparisons between the charge stability of different Ge QWs, we have summarized the change noise levels reported from different articles in Table 1. The Ge QWs with thickness of $15-20\,\mathrm{nm}$ are grown on Si or Ge substrates, and the thickness of the SiGe barriers varies from 4 to $55\,\mathrm{nm}$, while Al$_{2}$O$_{3}$, HfO$_{2}$ or SiO$_{2}$ is used as dielectric layers in quantum dot devices. Despite subtle differences in material structures and/or dielectrics, a quantitative comparison can reflect the electrical stability across different material structures. The last two columns are the reported charge noise amplitude $\sqrt{S_{\epsilon}}$ of quantum dot devices made from these materials at the frequencies of $1\,\mathrm{Hz}$ and $10\,\mathrm{mHz}$. The charge noise amplitudes are extracted via the two methods (current spectroscopy and Coulomb peak tracking), which are utilized in this work and yield consistent results over a wide frequency range (see Supplementary Fig. S12). As seen in the chart, the charge noise levels of our quantum dot devices are lower than that reported in other Ge QWs except for the most recent work\cite{stehouwer2025exploiting}, which reports slightly better performance than ours. Possible reasons for the lower charge noise in Ref. \zcite{stehouwer2025exploiting} include: (1) the use of Ge wafers as substrates, which reduces lattice mismatch between the substrate and the epitaxial layers; (2) a thicker ($55\,\mathrm{nm}$) SiGe barrier layer atop the Ge QW layer, which mitigates surface scattering. In principle, both techniques can be applied to CM SiGe/Ge QW heterostructures for further improvements. Nonetheless, the fact that introducing the composition modulation technique improves the charge stability of SiGe/Ge QW heterostructures to a level comparable with the state-of-the-art result underscores the capability of this approach. 

% For our data at $10\,\mathrm{mHz}$, the charge noise amplitude is extracted using two methods, yielding values that deviate reasonably from each other (see Figure S12 in the Supplementary).

% We find that the charge noise amplitudes $\sqrt{S_{\epsilon, f=10\,\mathrm{mHz}}}$ in our work get close to that in Ref.\cite{stehouwer2025exploiting}, indicating the remarkable charge stability of our quantum dot devices under prolonged operations.

% As a further demonstration of the charge stability of our devices over extended periods, Figure \ref{fig6} presents the charge stability of the double quantum dot in Device A. In the Figure \ref{fig6}a, the transconductance of charge sensor d$I_\mathrm{sd}$/d$V_\mathrm{P1}$ is measured as a function of the two plunger gate voltages. The numbers in brackets mark the hole numbers in the two quantum dots and different charge states are separated by clear boundaries. Repeated measurements of filling one hole into the right and left dots are performed by scanning corresponding plunger gates (black and white bars), and the results are presented in Figure \ref{fig6}b and \ref{fig6}c. The gate voltages of the charge transitions between (0, 0)-(1, 0) or (0, 0)-(0, 1) keep stable over 100 minutes, confirming the excellent charge stability of our Ge quantum dot devices.

\section*{Conclusion}
In conclusion, we have employed band-structure engineering to achieve SiGe/Ge QW heterostructures with exceptionally low charge noise. Leveraging the atomic-scale precision of MBE, we have engineered a composition-modulated barrier layer that lower the valance band, energetically forbidding charge accumulation at the SiGe-dielectric interface and thereby improving the charge stability. Hall device measurements reveal that the composition-modulated barrier helps expand the stable gate voltage range nearly threefold. Quantum dot devices made from the composition-modulated SiGe/Ge QW heterostructure exhibit remarkable charge stability in prolonged time periods. A quantitative comparison on the charge noise levels across different Ge QW materials validates the exceptional charge stability of our Ge quantum dot devices. These results demonstrate the power of band-structure engineering in low-noise Ge QWs and highlight the great potential of composition-modulated Ge/Si heterostructures for achieving advanced quantum devices.

\section*{Methods}
\subsection*{Band simulations}
For the numerical simulations, we employ a self-consistent scheme that solves the 1D Schrödinger and Poisson equations iteratively until both the band alignment and charge distribution converge. The simulations are performed within a python frame revised from an open-source package named “Aestimo”\cite{hebal2021general}. A three-point finite difference scheme is introduced to solve the Schrödinger equation with varied effective mass \cite{Tan1990A}, where the matrix elements of the kinetic energy part of the Hamiltonian can be approximated as:
% Requires: \usepackage{amsmath}
\begin{equation}
    \label{eq:placeholder_label}
    \frac{d}{dx}\left(\frac{1}{m^\ast(x)}\frac{d\psi}{dx}\right)
    \approx \frac{1}{\Delta x^2}
    \left[
        \frac{\psi_{i+1}-\psi_i}{m^\ast_{i+1/2}}
        - 
        \frac{\psi_i-\psi_{i-1}}{m^\ast_{i-1/2}}
    \right].
\end{equation}
The effective masses of the SiGe layers are obtained via linear interpolation of Si and Ge values, using a Ge hole effective mass of 0.08 m\textsubscript{e}\cite{zhang2024high} and a Si value of 0.23 m\textsubscript{e}\cite{sant2013band}. The dielectric constant of the SiGe layers is determined via linear interpolation between the values for pure Ge (16.0) and pure Si (11.7) depending on local compositions. The valence band alignment of the heterostructures is extracted from the parameters of strained SiGe layers \cite{sant2013band}. To avoid numerical divergence arising from “floating-point underflow”, the temperature is set to 30 K in the solution of the Poisson equation. Furthermore, into the computational model are incorporated the following approximations and simplifications: (1) For the CM SiGe/Ge QW heterostructures, the Ge concentration in the SiGe layers is modeled with a linear gradient with respect to depth. (2) The interfaces of SiGe/Ge QW heterostructures are modeled as perfectly abrupt, manifesting as sharp discontinuities in the valence band maximum. 

\subsection*{SiGe/Ge QW heterostructure growth}
The heterostructures are grown in a 4-inch-wafer MBE system with a background vacuum of $3\times10^\mathrm{-10}\,\mathrm{mbar}$. The employed Si and Ge solid sources is 99.99999\% and 99.9999\% in purity, respectively. Commercial electron beam evaporators are used to evaporate Si and Ge during the material growth. The SiGe/Ge QW heterostructures (H0 and H1) in this work are grown on Si$_{0.2}$Ge$_{0.8}$ virtual substrates, which are epitaxially grown on a Si(001) wafer via a reverse-graded-buffer technique (Supplementary Section I). Heterostructure H1 is grown as follows. (1) A $400\,\mathrm{nm}$ thick layer of Si$_{0.2}$Ge$_{0.8}$ is initially grown at 450 \textsuperscript{\textsuperscript{o}}C with a rate of 1 Å/s to serve as the bottom barriers. (2) Subsequently, a 15-nm-thick Ge QW layer is grown at 450 \textsuperscript{\textsuperscript{o}}C with a growth rate of 0.8 Å/s. (3) On top of the Ge layer, the deposition of the SiGe top barrier is done in two steps. First, a 10-nm-thick Si$_{0.2}$Ge$_{0.8}$ spacer barrier layer is grown at 450 \textsuperscript{\textsuperscript{o}}C with a growth rate of 1 Å/s. Second, a 25-nm-thick CM barrier layer is deposited at 390 \textsuperscript{\textsuperscript{o}}C, with its composition gradually tuned from Si$_{0.2}$Ge$_{0.8}$ at the bottom to Si$_{0.8}$Ge$_{0.2}$ at the top. The modulations of compositions are achieved by independently adjusting the deposition rates of Ge and Si. The Ge deposition rate is varied from 0.8 Å/s to 0.1 Å/s during the growth, while the Si rate is accordingly adjusted from 0.2 Å/s to 0.4 Å/s. (4) As a final step, a 1-nm-thick Si capping layer is deposited on the top at 330 \textsuperscript{\textsuperscript{o}}C. In contrast to heterostructure H1, the growth of H0 differs in the top barrier, where 35-nm-thick Si$_{0.2}$Ge$_{0.8}$ without composition modulation is grown at 450 \textsuperscript{\textsuperscript{o}}C with a growth rate of 1 Å/s.

\subsection*{Device fabrications}

In this work, Hall devices are made from both two SiGe/Ge QW heterostructures (H0 and H1), and quantum dot devices are made from heterostructure H1. The fabrication of both Hall devices and quantum dot devices includes mesa etching, ohmic contact formation, dielectric layer deposition, and top gate definition. For the Hall devices, the fabrication begins with mesa etching by Reactive Ion Etching (RIE). After that, a diluted buffered oxide etch (BOE) solution is used to remove the oxidation layer in the contact region, prior to the deposition of 60-nm-thick Pt. An annealing step at 300 \textsuperscript{\textsuperscript{o}}C for 2 hours is performed to diffuse Pt into the quantum well to achieve ohmic contacts. Then, 25-nm-thick $\mathrm{Al}_2\mathrm{O}_3$ is deposited by atomic layer deposition (ALD) as the dielectric layer. As a final step, the top gate is made by depositing $10/150\,\mathrm{nm}$ Ti/Au. In contrast to the Hall devices, the fabrication of the quantum dot devices differs in contact metal thickness ($30\,\mathrm{nm}$ Pt), $\mathrm{Al}_2\mathrm{O}_3$ thickness ($10\,\mathrm{nm}$) and gate electrodes ($3/27\,\mathrm{nm}$ Ti/Pd).

\subsection*{Hall device measurements}

The Hall devices are characterized in a cryostat with a base temperature of $2\,\mathrm{K}$. Electrical measurements are conducted using a standard lock-in technique at a frequency of $17.777\,\mathrm{Hz}$. An AC voltage $V_\mathrm{ac}=1\,\mathrm{V}$ is applied across a $10\,\mathrm{M\Omega}$ resistor connected to the source terminal, while the drain terminal is connected to an AC current meter. This configuration ensures a stable AC current of $100\,\mathrm{nA}$ when the device channel is turned on. The top-gate voltage, $V_\mathrm{g}$, is applied via a DC voltage source to modulate the carrier concentration within the channel.

% The quantum transport properties of 2DHG in the C-M QW material are investigated by PPMS at 1.7 K, with details presented in Supporting Information. A pronounced integer quantum Hall effect is observed. Compared to MBE SiGe/Ge/SiGe QWs with a uniform barrier composition, the gate-tunable range of the 2DHG hole concentration in the C-M QW is enhanced by a factor of two. A 2DHG quantum lifetime of 1.43 ps is achieved at 1.7 K, reflecting the advantage of the C-M structure in reducing remote impurity scattering (see Supporting Information)\cite{zhang2024high}. The hole effective mass (m*) and effective g-factor (g*) were proved to be gate-tunable from 0.122 m\textsubscript{e} to 0.077 m\textsubscript{e} and from 4.5 to 9.7, respectively, demonstrating an advanced gate tunability of hole properties in the C-M QW. 

Figure \ref{fig3}c shows the dependence of $V_\mathrm{t.o.}$ on $V_\mathrm{sn}$ for Hall devices. Here, the threshold voltage $V_\mathrm{t.o.}$ is defined as the gate voltage at which the device current reaches 90\% of its saturation value in a transfer characteristic curve. The dashed lines in Fig. \ref{fig3}c show the linear fits to the $V_\mathrm{t.o.}-V_\mathrm{sn}$ curves in the linear shift zones, with a function of $V_\mathrm{t.o.}=k\times V_\mathrm{sn}-b$. The intersection of these dashed lines with the $V_\mathrm{t.o.}$ in the stable zone defines the critical voltage $V_\mathrm{sn}^{*}$. The stable voltage range, $V_\mathrm{stable}$, is defined by the interval between $V_\mathrm{sn}^{*}$ and the initial voltage $V_\mathrm{sn}^{0}$.

% A coefficient of \textit{\textit{V}}\textsubscript{stable} is introduced to evaluate the voltage stability of Hall device, defined as 
% \(\textit{\textit{V}}\textsubscript{stable}=\textit{\textit{V$^{*}$}}\textsubscript{t.o. } -(\textit{\textit{V$^{*}$}}\textsubscript{t.o. } +b)/k\)
% , where \textit{\textit{V$^{*}$}}\textsubscript{t.o. } denotes the initial value of \textit{\textit{V}}\textsubscript{t.o. }measured after PPMS cooling but before charge trapping. Note that \textit{\textit{V}}\textsubscript{stable} equals the lateral voltage difference from \textit{\textit{V$^{*}$}}\textsubscript{t.o. } to the fit line. Therefore, \textit{\textit{V}}\textsubscript{stable} increases with the expansion of the stable zone in \textit{\textit{V}}\textsubscript{t.o.}-\textit{\textit{V}}\textsubscript{sn} curves. \textcolor{red}{The value of \textit{\textit{V}}\textsubscript{stable} is found to be an intrinsic material characteristic and independent of device manufacturing fluctuations (details can be found in \textbf{\textbf{Supporting Information}}). }

\subsection*{Charge noise characterizations}
\noindent\textbf{Current spectroscopy method}
% We set a source-drain bias about $0.1\,\mathrm{mV}$ using YOKOGAWA GS200 in quantum dot devices, applying positive voltage into barrier gates and plunger gate to tune the electrochemical potential of quantum dot until measured Coulomb oscillation. And then we measure the traces of source-drain current $I_{sd}$ on the Coulomb peak flank where the maximum slope is detected by QDAC-II (a 24-channel high-precision low-noise voltage generator for DC) and DMM of Keysight 34461A at the rate of $20\,\mathrm{ms}$ and lasted $1400\,\mathrm{s}$.

In Fig. \ref{fig4}, the current traces are recorded over time with a sampling interval of $20\,\mathrm{ms}$. Each trace is divided into eight segments; the power spectral density, $S_\mathrm{I}$, is then obtained by computing the Fast Fourier Transform (FFT) of each segment and averaging the results. The noise power spectral density $S_\mathrm{\epsilon}$ can be obtained from $S_\mathrm{I}$ via the relationship,
\begin{equation}
S_\mathrm{\epsilon} = \frac{\alpha^2 S_\mathrm{I}}{\left| dI_\mathrm{sd}/dV_\mathrm{P} \right|^2},
\end{equation}
where $\alpha$ is the lever arm of corresponding peaks and $dI_\mathrm{sd}/dV_\mathrm{P}$ is the maximum slope of the transport peak. These relevant parameters for Fig. \ref{fig4} are extracted from Supplementary Fig. S8.

\noindent\textbf{Coulomb peak tracking method}

In Fig. \ref{fig5}, we track the Coulomb peaks by repeatedly scanning $V_\mathrm{P}$. To precisely determine the peak position of each scan, the data are fitted with a hyperbolic secant function
\begin{equation}
f(x) = \frac{A}{cosh^2[B\times(x-x_0)]}+C,
\end{equation}
where $f(x)$ is the measured current, $x$ and $x_\mathrm{0}$ are the plunger gate voltage and the voltage of the peak center, respectively, and A, B and C are parameters for fitting. We separate the voltage fluctuations for four segments and calculate voltage power spectral density $S_\mathrm{V}$ with FFT. Corresponding charge noise spectral density $S_\mathrm{\epsilon}$ is obtained as  
\begin{equation}
S_\epsilon = \alpha^2 S_V.
\end{equation}
The lever arm, $\alpha$, for Fig. \ref{fig5} are extracted from Supplementary Fig. S10.
% \noindent\textbf{Voltage PSD of occupying single hole in charge sensor}

% We tune gates of charge sensor dot until observe Coulomb oscillation and set plunger gate of charge sensor dot to flank because it sensitive to charge fluctuation in environment (see more details in Figure S8 of Supplementary Materials), at the same time, we sweep plunger gates of DQD of $V_{P1}$ and $V_{P2}$ to few-hole regime. We record current of charge sensor dot when the first hole is loaded in quantum dot repeatedly by plunger gate voltage as Figure \ref{fig5}. A sigmoid function is used for fitting the track for each sweep
% \begin{equation}
%     f(x) = \frac{a}{1+e^{(x-x_0)/\tau}}+b
% \end{equation}
% where $f(x)$ and $x_0$ is the measured current and centered voltage which load into the first hole of quantum dot, respectively, and $a$, $b$ and $\tau$ are parameters for fitting. At last, we calculate voltage PSD. 

\section*{Data availability}

The raw data generated in this study, as well as the code used to analyze the data can be obtained from corresponding authors upon reasonable request.

\section*{Acknowledgments}
D.-M.H. and J.-H.L. contributed equally to this paper. We would like to thank Xiao-Bo Li for his assistance in Hall device fabrications. This work was financially supported by the National Key Research and Development Program of China (No.2025YFE0217400), the NSFC (Nos. 92565304, 62225407, 92576112, 12374480, 92576112, 12304207, 12304101, 12304100, 92165207, 92165208, 62501059), Beijing Natural Science Foundation (No.JQ26005) and the Quantum Science and Technology-National Science and Technology Major Project (No.2021ZD0302300).

\section*{Author contributions}

H.Q.X. conceived and supervised the project. D.-M.H. designed the layer structures of the materials and ran the band-structure simulations. D.-M.H., J.-Y.Z., J.-H.W. and X.-Y.Z. conducted the material growth under the supervision of J.-J.Z.. F.-Z.L., D.-M.H., J.-H.W. and B.-X.F. contributed to the material characterizations. J.-H.L. conducted device fabrications. D.-M.H. and H.G. carried out the Hall measurements, and J.-H.L., J.-Y.W. and Y.L. performed the measurements on the quantum dot devices. J.-H.L., D.-M.H., X.-F.L., J.-Y.W., J.-J.Z. and H.Q.X. performed the data analysis. D.-M.H., J.-Y.W., J.-H.L. and H.Q.X. prepared the manuscript with comments from the other co-authors. 

\section*{Competing interests}
The authors declare no competing interests.

\bibliographystyle{science}
% \bibliography{refs}

\clearpage

\end{document}

\baselineskip24pt
\maketitle 

\noindent \textbf{Section I. Virtual substrate growth}--Fig. S1

\noindent \textbf{Section II. Characterization of material composition in H1}--Fig. S2

\noindent \textbf{Section III. Material characterizations of H0}--Fig. S3

\noindent \textbf{Section IV. Hall measurements of SiGe/Ge QW heterostructures}--Figs. S4-S6

\noindent \textbf{Section V. Data of quantum dot devices}--Figs. S7-S12

\newpage

\section*{I. Virtual substrate growth}

\begin{figure}[!b] 
\centering
\includegraphics [width=1\linewidth]{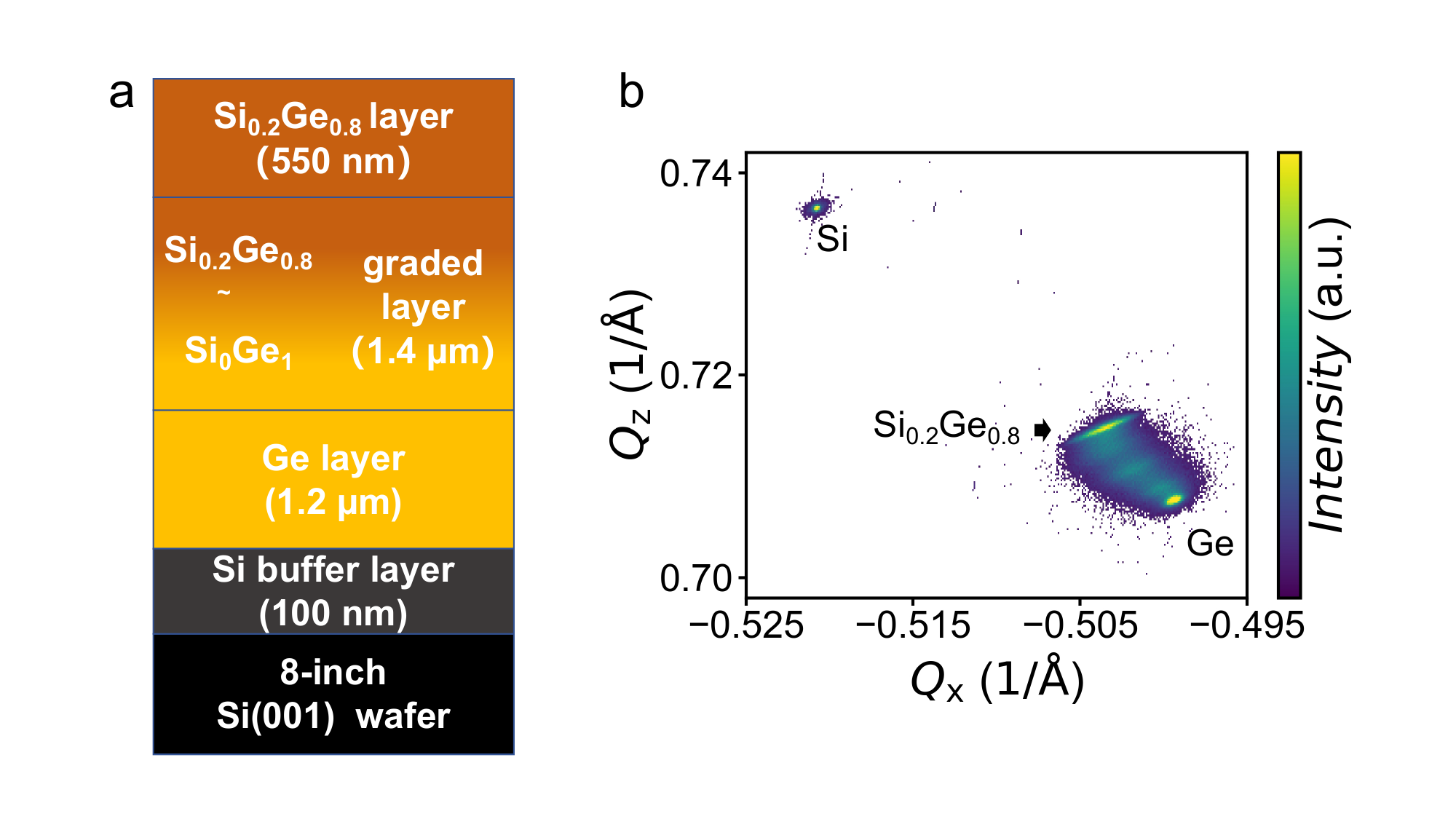}
\caption{\textbf{Structure of the virtual substrate.} \textbf{a}, Schematic of the epitaxial layers. \textbf{b}, XRD-RSM of the virtual substrate. Three main intensity peaks arising from Si, Si$_\mathrm{0.2}$Ge$_\mathrm{0.8}$ and Ge are observed. The signal between the Si$_\mathrm{0.2}$Ge$_\mathrm{0.8}$ peak and the Ge peak arises from the strain-relaxed, composition-graded SiGe layer.}
\end{figure}
%\clearpage

The virtual substrates were grown on 8-inch Si (001) wafers using a reverse compositional grading approach via molecular beam epitaxy (MBE). Prior to the growth, the substrates underwent a cleaning procedure to remove native oxides, involving immersion in a diluted hydrofluoric acid solution (HF:H$_\mathrm{2}$O = 1:10) followed by rinsing with deionized water (resistivity $\sim 18\,\mathrm{M\Omega \cdot cm}$). Subsequently, the substrates were loaded into the MBE chamber for degassing and dehydrogenation. Figure S1a shows a schematic of the virtual substrate epitaxial layer structure. The epitaxial growth process started with the deposition of a 100-nm-thick Si buffer layer at 400 °C with a growth rate of 1 Å/s. Subsequently, a 1.2-$\mu$m-thick Ge layer was grown through eight cycles of alternating Ge deposition at 280 °C and \textit{in situ} annealing at 720 °C. A 1.4-$\mu$m-thick linearly graded Si$_\mathrm{x}$Ge$_\mathrm{1-x}$ layer was then grown by adjusting the relative deposition rates of Si and Ge during growth, with the composition gradually changing from pure Ge (bottom) to Si$_\mathrm{0.2}$Ge$_\mathrm{0.8}$ (top). Finally, a 550-nm-thick Si$_\mathrm{0.2}$Ge$_\mathrm{0.8}$ layer was deposited to ensure a high quality surface. This optimized growth strategy ensures high-quality strain-relaxed Si$_\mathrm{0.2}$Ge$_\mathrm{0.8}$ virtual substrates. Figure S1b shows the XRD-RSM of the virtual substrate, calibrated via the left-top intensity peak of the Si (\(\bar{2}\bar{2}4\)) facet. Three main intensity peaks corresponding to Si, Si$_\mathrm{0.2}$Ge$_\mathrm{0.8}$ and Ge are observed. The signal between the Ge peak and the Si$_\mathrm{0.2}$Ge$_\mathrm{0.8}$ peak arises from the strain-relaxed, graded SiGe layer. 

\section*{II. Characterization of material composition in H1}

\begin{figure}[!b] 
\centering
\includegraphics [width=0.75\linewidth]{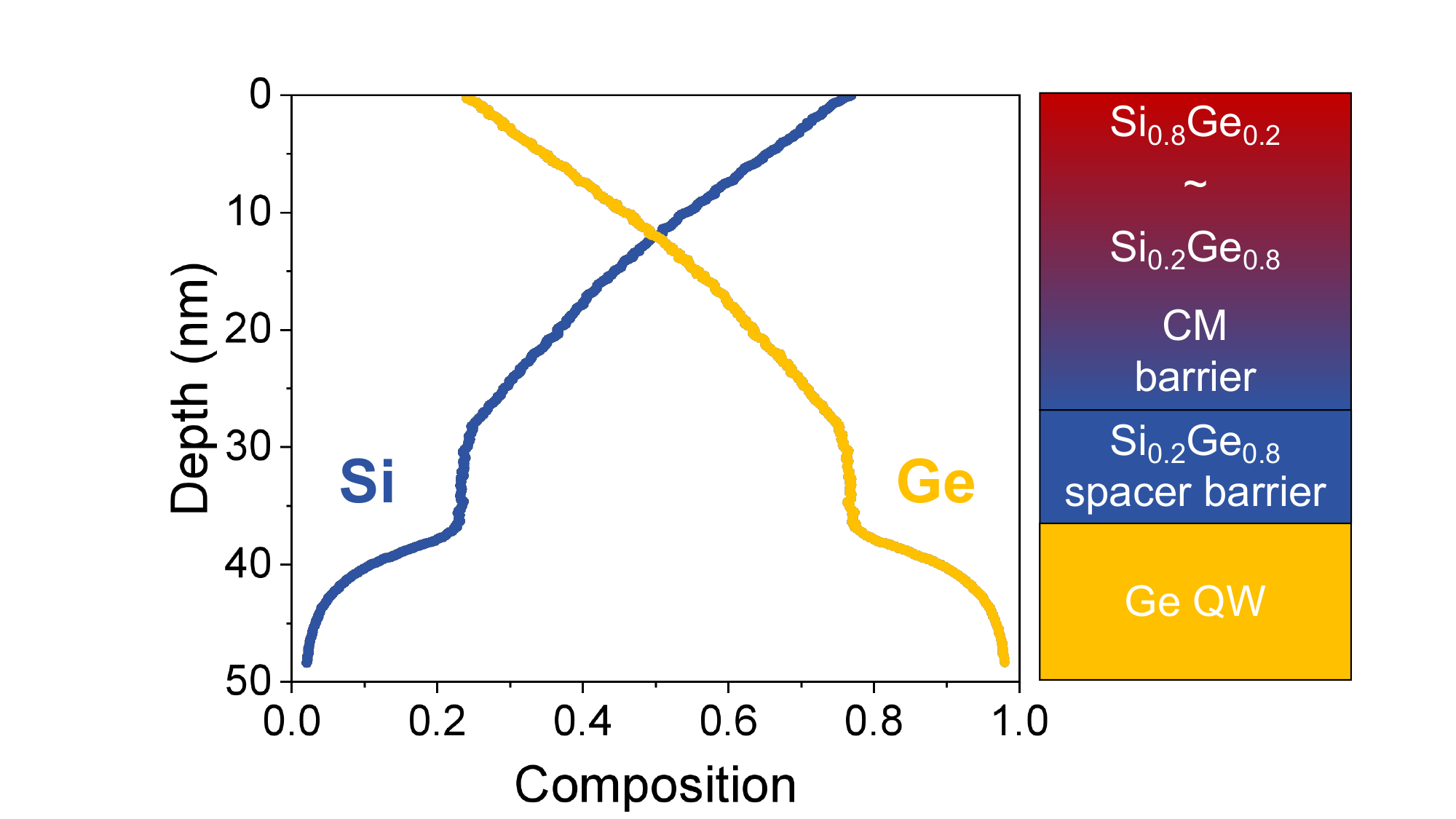}
\caption{Depth profile of the composition-modulated SiGe/Ge quantum well heterostructure (H1) measured via SIMS (left panel) and epitaxial layer structure of the material (right panel).}
\end{figure}

The measured compositions of the epitaxial layers show discrepancies across different characterization techniques, attributed to variations in analytical accuracy, sampling volumes and calibration standards. Energy-dispersive spectroscopy (EDS) data indicate a Si:Ge ratio of 0.17:0.83 in the spacer barrier layer of H1 (Fig. 2b). In contrast, X-ray diffraction (XRD) analysis yields a ratio of 0.21:0.79 (Fig. 2c). Figure S2 shows the depth profile of secondary ion mass spectrometry (SIMS), revealing a distinct ratio of 0.23:0.77 in the spacer barrier layer. These systematic variations suggest that the actual composition lies within the range bounded by these measurements. Although there is an absolute composition uncertainty of ±4\%, all three techniques consistently demonstrate the same compositional gradient trend in the top barrier. The averaged result of composition in the spacer barrier layer (10-nm-thick Si$_\mathrm{0.2}$Ge$_\mathrm{0.8}$ layer) perfectly matches the value derived by the deposition rate. 

%\clearpage
\section*{III. Material characterizations of H0}
\begin{figure}[!h] 
\centering
\includegraphics [width=1\linewidth]{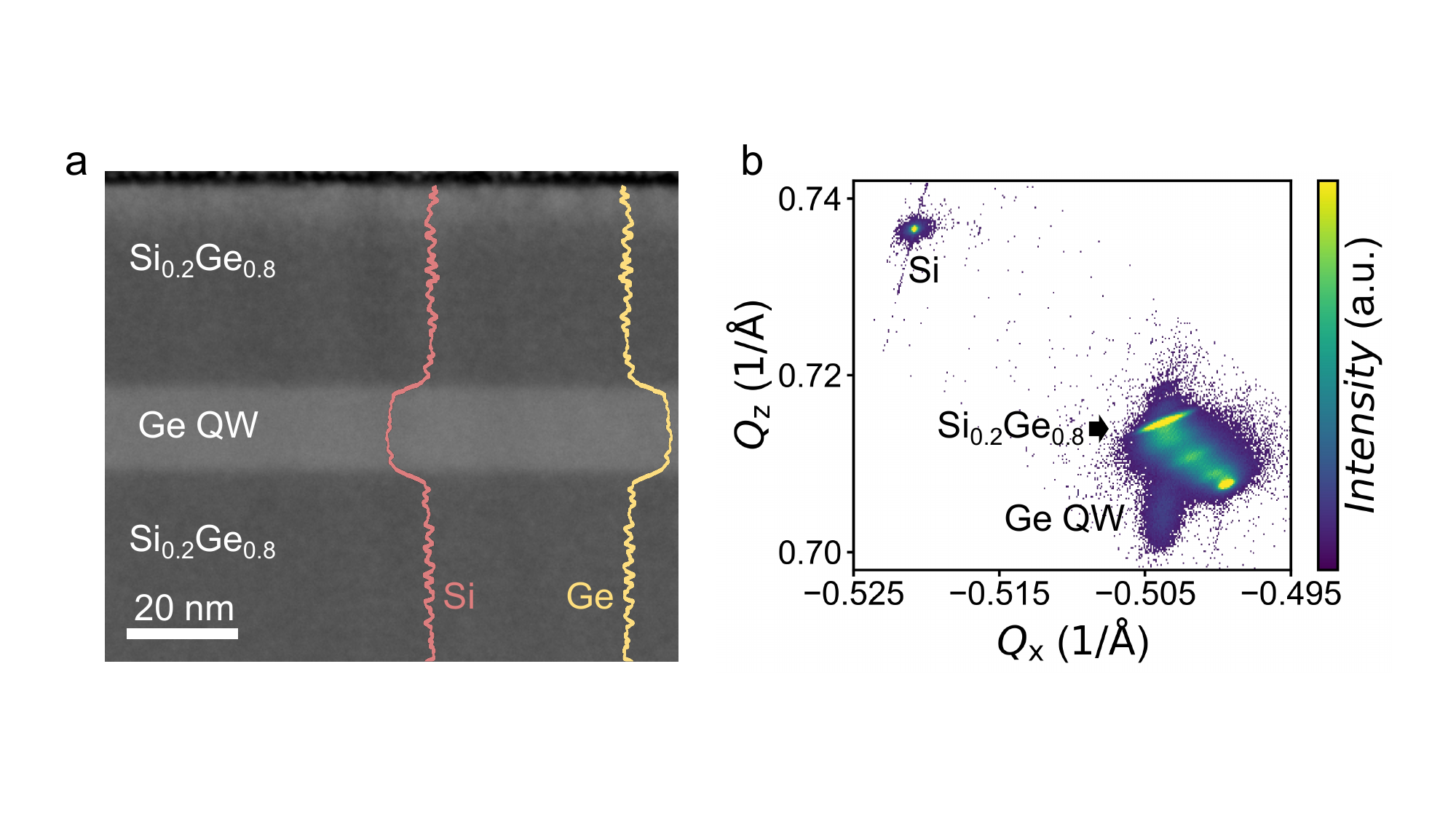}
\caption{\textbf{Basic material characterizations of the reference heterostructure H0.} \textbf{a}, HAADF TEM image of the SiGe/Ge QW region. The EDS profiles of Ge (yellow) and Si (pink) are overlaid in the panel. \textbf{b}, XRD reciprocal space mapping of H0. The in-plane lattice constant of the Ge QW layer matches that of the Si\textsubscript{0.8}Ge\textsubscript{0.2} layers, indicating a fully strained heterostructure.}
\end{figure}

%\clearpage

\section*{IV. Hall measurements of SiGe/Ge QW heterostructures}

Four Hall devices made from heterostructure H0 (H0-Dev1 to H0-Dev4) and five Hall devices made from heterostructure H1 (H1-Dev1 to H1-Dev5) are characterized in electrical transport measurements. Figures S4 and S5 present the electrical measurement results from these Hall devices, and Fig. S6 presents simulated band structures and hole density distributions of the two heterostructures.

\begin{figure}[!b] 
\centering
\includegraphics[width=\linewidth] {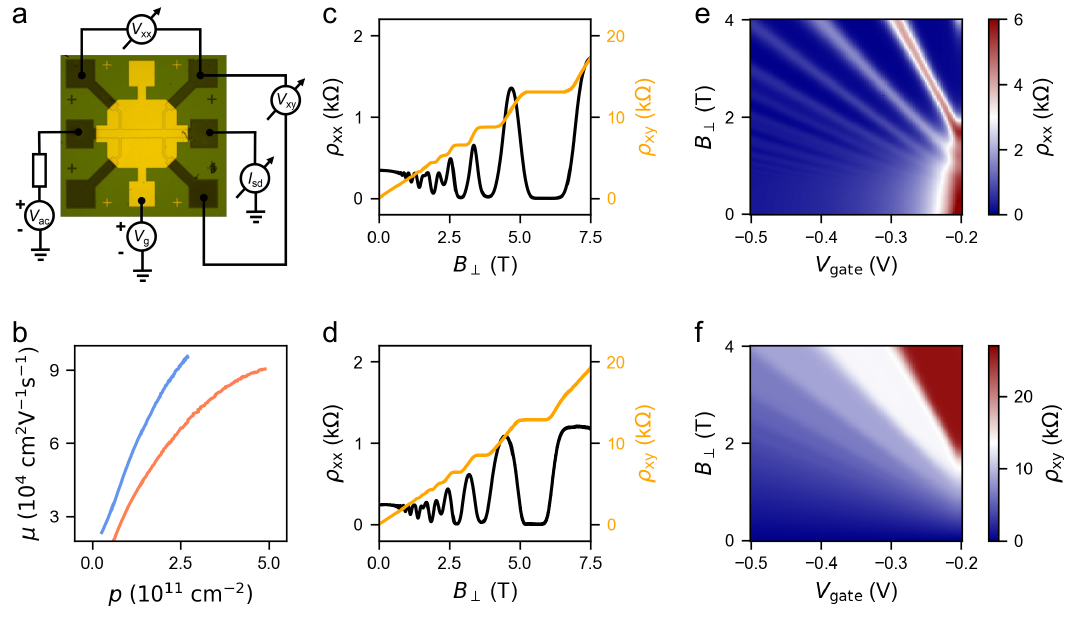}
\caption{\textbf{Characterizations of the Hall devices made from H0 and H1.} \textbf{a}, Optical microscope image of a Hall device with a typical measurement circuit setup. \textbf{b}, Measured hole mobility as a function of hole concentration for H0-Dev1 (blue) and H1-Dev2 (orange). The maximum hole mobility for both devices is $\sim90,000\,\mathrm{cm^{2}/V\cdot s}$. At an identical hole concentration, H1-Dev2 shows a slightly lower mobility, likely due to strain from compositional modulation. \textbf{c}, $\rho$$_\mathrm{xx}$ and $\rho$$_\mathrm{xy}$ as functions of magnetic field for H1-Dev2, measured at a hole concentration of 2.7 × 10$^\mathrm{11}\,\mathrm{cm^{-2}}$. The discrete peaks in $\rho$\textsubscript{xx} and quantized plateaus in $\rho$\textsubscript{xy} indicate a pronounced integer quantum Hall effect. \textbf{d}, $\rho$$_\mathrm{xx}$ and $\rho$$_\mathrm{xy}$ as functions of magnetic field for H0-Dev1, measured at a hole concentration of 2.7 × 10$^\mathrm{11}\,\mathrm{cm^{-2}}$. \textbf{e}, Landau-fan diagram of the longitudinal resistance $\rho$$_\mathrm{xx}$ for H1-Dev5. \textbf{f}, Landau-fan diagram of the transverse resistance $\rho$$_\mathrm{xy}$ for H1-Dev5.}
\end{figure}

% \begin{figure}[!t] 
% \centering
% \includegraphics[width=\linewidth] {FigSI5}
% \caption{\textbf{Temperature dependent SdH oscillations.} \textbf{a.} Temperature dependence of the re-calibrated SdH oscillations $\Delta$$\rho$\textsubscript{xx} at a hole concentration of 5.23 × 10\textsuperscript{11 } cm\textsuperscript{-2}. \textbf{b.} $\Delta$$\rho$\textsubscript{xx} as a function of temperature at a magnetic filed of 1.55 T from \textbf{a}. The red curve shows the fitting for extracting \textit{m}*. \textbf{c.} Temperature dependent $\Delta$$\rho$\textsubscript{xx} at a hole concentration of 2.75 × 10\textsuperscript{11} cm\textsuperscript{-2}. \textbf{d.} $\Delta$$\rho$\textsubscript{xx} as a function of temperature at 1.3 T from \textbf{c}. \textbf{e-f.} Raw data of SdH oscillations at hole concentrations of 5.23 × 10\textsuperscript{11} cm\textsuperscript{-2} and 2.75 × 10\textsuperscript{11} cm\textsuperscript{-2}, respectively. }\label{fig:AlOx_rdrain}
% \end{figure}

% \clearpage

% \section*{V. \textbf{\textit{\textit{V}}}\textbf{\textsubscript{t.o.}}\textbf{-}\textbf{\textit{V}}\textbf{\textsubscript{sn}}\textbf{ curves across multiple Hall devices}}

\begin{figure}[!b] 
\centering
\includegraphics[width=\linewidth] {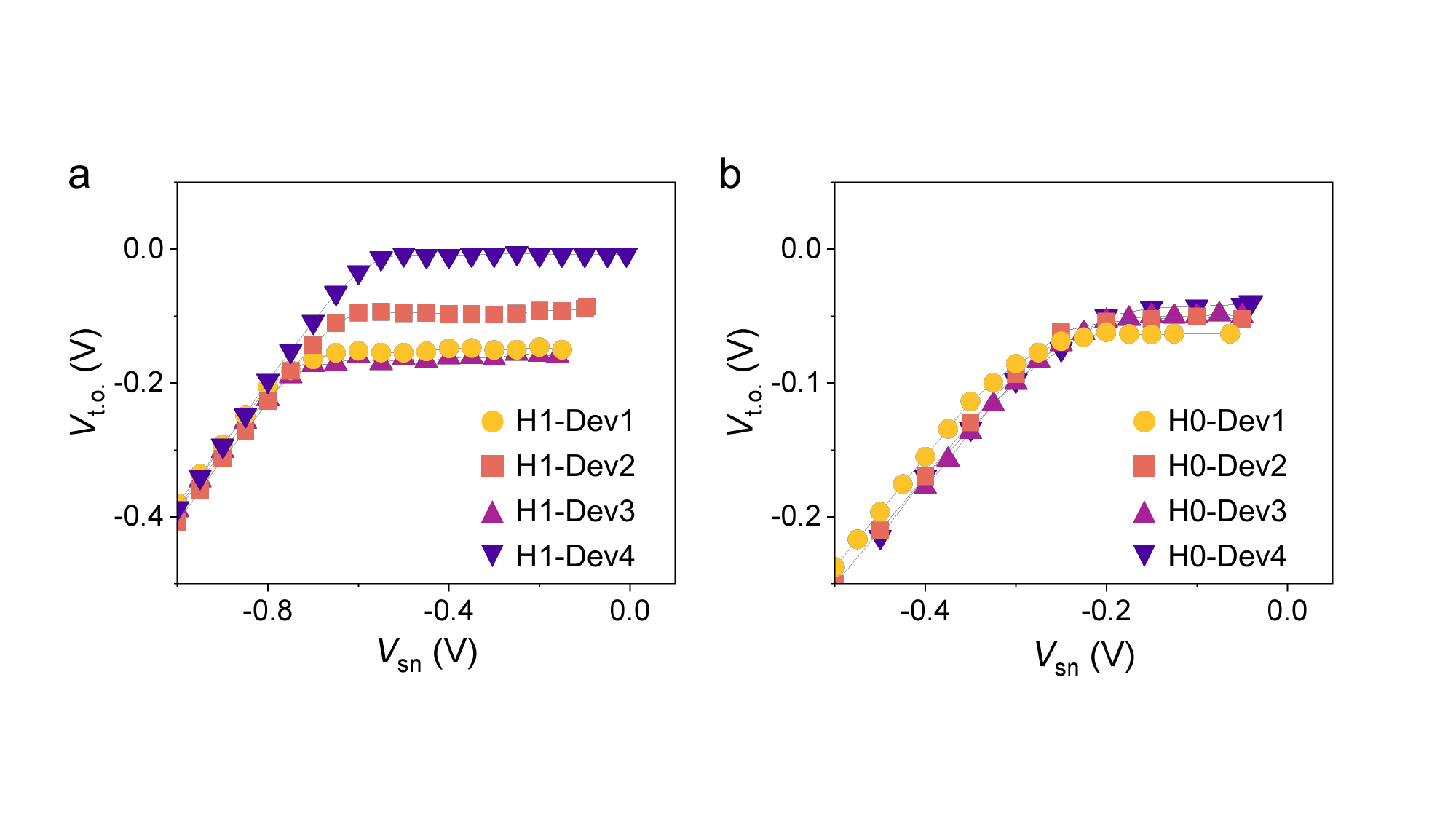}
\caption{\textbf{Raw experimental data of $V_\mathrm{t.o.}$-$V_\mathrm{sn}$ curves across various Hall devices.} \textbf{a}, $V_\mathrm{t.o.}$-$V_\mathrm{sn}$ relations for four devices made from H1. \textbf{b}, $V_\mathrm{t.o.}$-$V_\mathrm{sn}$ relations for four devices based on H0. The initial threshold voltage of these 8 devices exhibits fluctuations, resulting from device fabrication or material inhomogeneities. The parameter $V_\mathrm{stable}$ derived from these curves correlates with material structures and is almost independent of device fluctuations.}
\end{figure}
\clearpage

\begin{figure}[!b] 
\centering
\includegraphics[width=\linewidth] {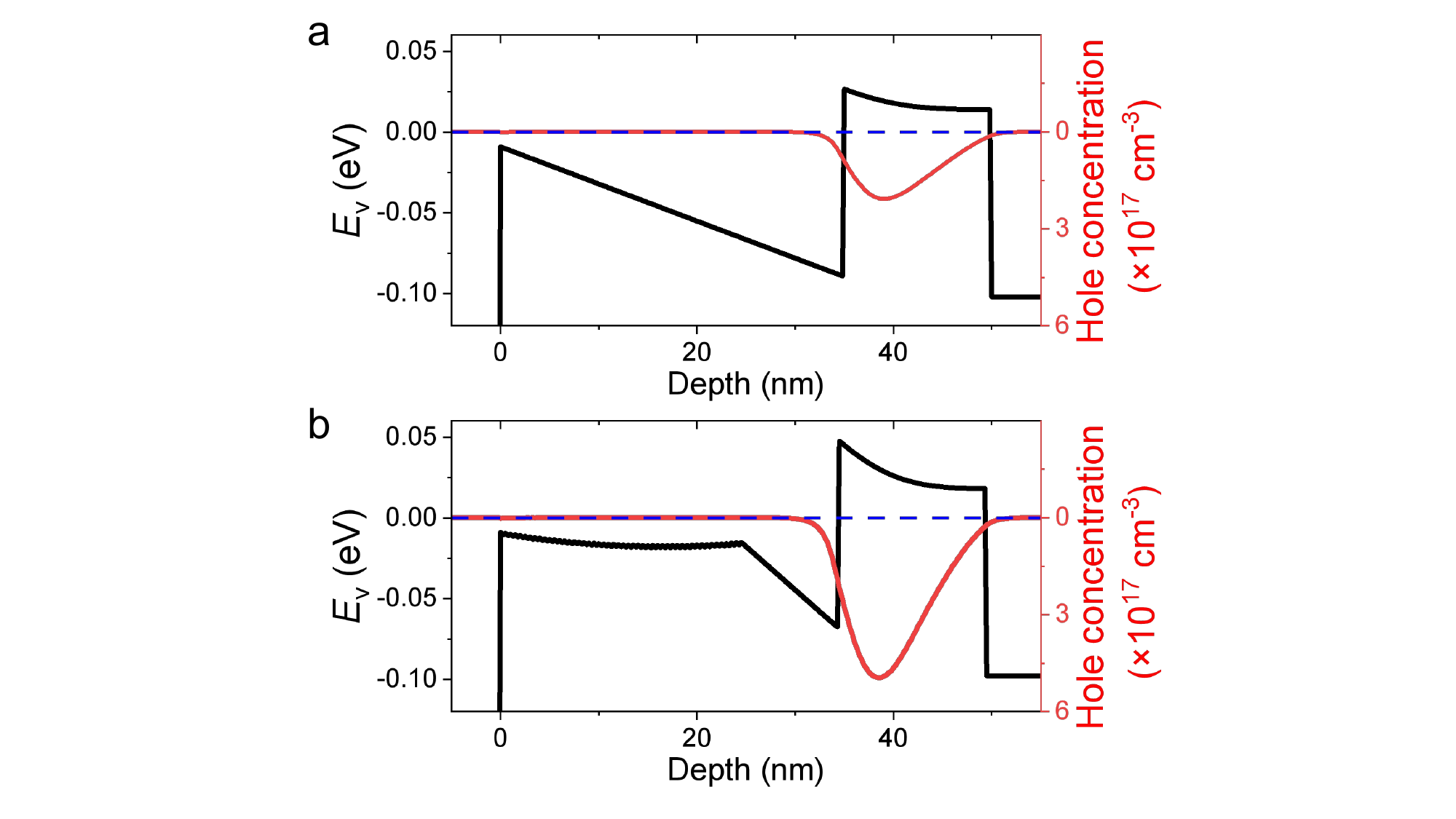}
\caption{\textbf{Simulated band structures and hole density distributions of heterostructures H0 and H1.} \textbf{a}, Valence band diagram and hole distributions of H0 at a critical gate voltage, $V_\mathrm{gate}=-0.430\,\mathrm{V}$, below which charge trapping could occur (the same condition as that in Fig. 1c, where the valence band maximum at the SiGe-dielectric interface is only 10 meV below the Fermi level). \textbf{b}, Valence band diagram and hole density distributions of H1 at a critical gate voltage, $V_\mathrm{gate}=-0.633\,\mathrm{V}$, where the valence band maximum at the SiGe-dielectric interface is 10 meV below the Fermi level. Compared to H0, much higher hole concentration can be reached in H1 before charge trapping occurs.}
\end{figure}
\clearpage

\newpage
\section*{V. Data of quantum dot devices}

Three quantum dot devices (device A-C) are made from the composition-modulated SiGe/Ge QW heterostructure---H1. Figure S7 displays the scanning electron microscope (SEM) images of the three devices. Figures S8 and S9 present data in the current spectroscopy method. Figures S10 and S11 correspond to the method of Coulomb peak tracking. Figure S12 shows a comparison of the charge noise data extracted with the two methods.

\begin{figure}[!h] 
\centering
\includegraphics[width=\linewidth] {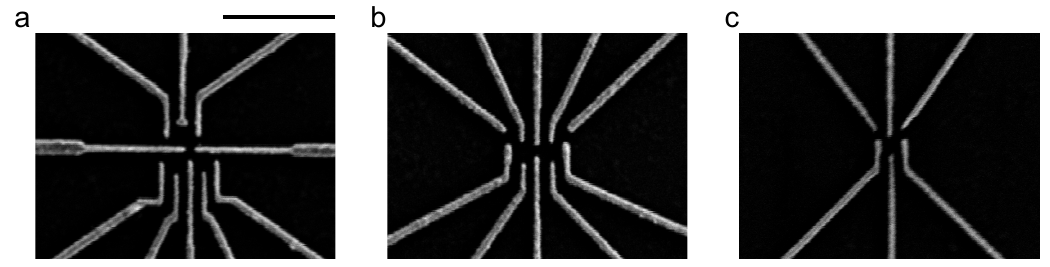}
\caption{\textbf{SEM images of quantum dot devices for charge noise spectra measurements.} \textbf{a}, Device A of a charge sensor dot and a double quantum dot. \textbf{b}, Device B of a double quantum dot. Only the right dot is used in this work. \textbf{c}, Device C of a single quantum dot. The scale bar is $500\,\mathrm{nm}$.}
\end{figure}

\begin{figure}[!b] 
\centering
\includegraphics[width=\linewidth] {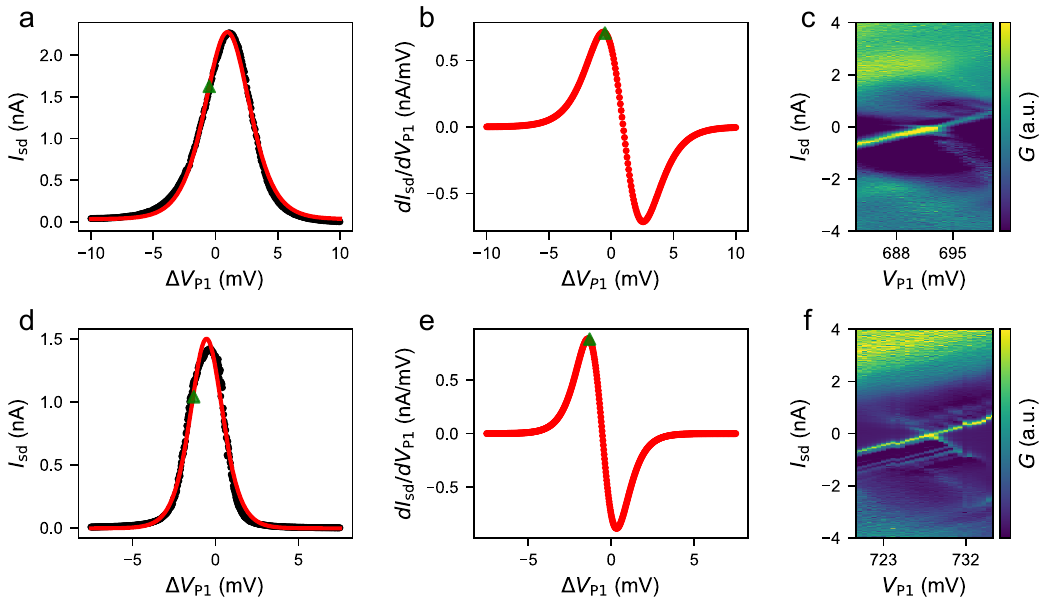}
\caption{\textbf{Key parameters for quantifying charge noise in current spectroscopy method.} \textbf{a}, Coulomb oscillation of Peak 1 in device A. Black points represent raw data, and the red curve is a fit using the hyperbolic secant function (see Methods). Current traces are extracted at the point of maximum transconductance, marked by a green triangle. \textbf{b}, Calculated transconductance from the fitting curve in figure a. The maximum value marked by a green triangle is used for converting $S_\mathrm{I}$ to charge noise power spectral density $S_\mathrm{\epsilon}$ (see Methods). \textbf{c}, Coulomb diamond of Peak 1 in device A for level arm extraction. The lever arm $\alpha$ is extracted by the formulation of $\alpha = \left| \frac{m_s \cdot m_d}{m_s - m_d} \right|$, where $m_s$ and $m_d$ are the slopes of Coulomb diamond boundaries when the source and drain are aligned in energy with the quantum dot level, respectively. \textbf{d-f}, Corresponding Coulomb oscillation, transconductance curve and Coulomb diamond for Peak 2 of device A.}
\end{figure}

\begin{figure}[!t] 
\centering
\includegraphics[width=\linewidth] {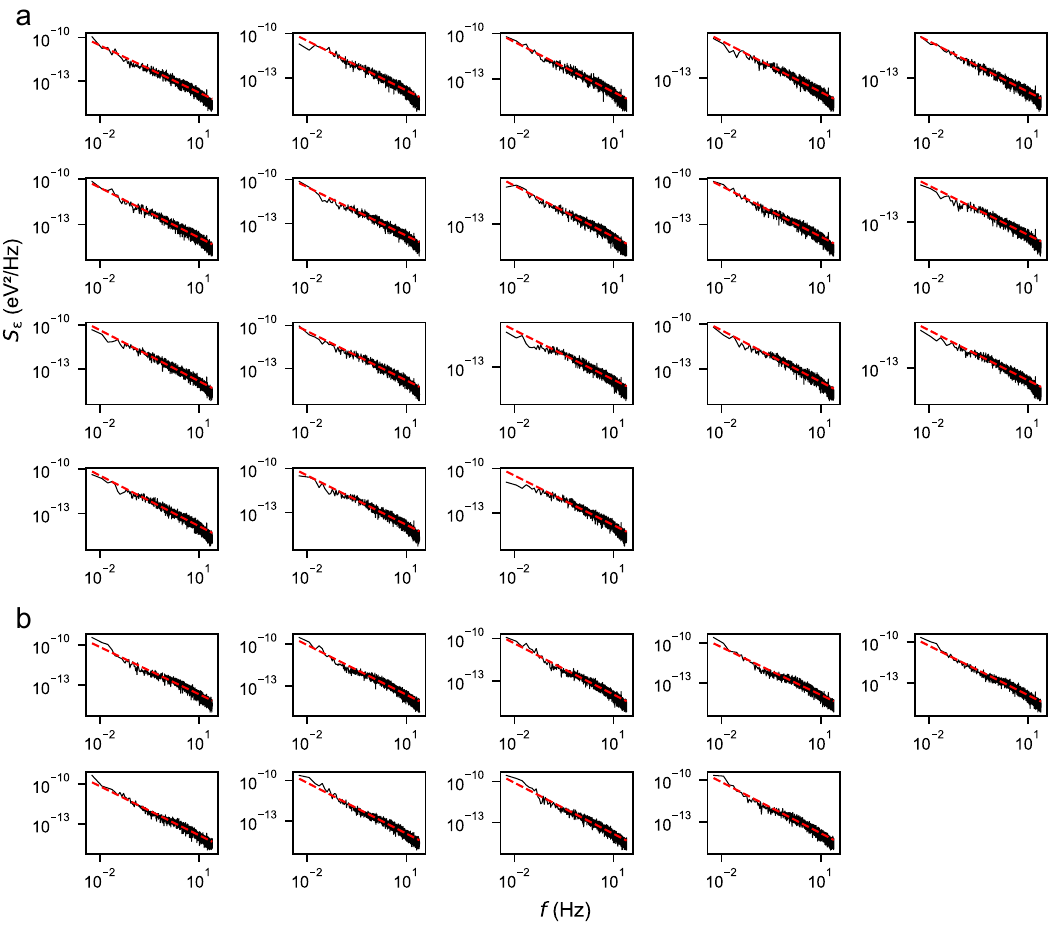}
\caption{\textbf{Charge noise spectra obtained via current spectroscopy.} \textbf{a} and \textbf{b}, Charge noise spectra of Peak 1 and Peak 2 in device A. The key parameters for converting $S_\mathrm{I}$ to $S_{\epsilon}$ are from Fig. S8. The red dashed lines are the fitting curves to the data via the formula of $1/f^{\alpha}$, with the fitting range of 5 mHz to 10 Hz. In fittings, $\alpha$ is between 1.05 and 1.25. The charge noise amplitudes $\sqrt{S_\epsilon}$ at $1\,\mathrm{Hz}$ and $10\,\mathrm{mHz}$ are extracted from the fitting curves, and presented in Figs. 4e and 4f. The average charge noise amplitude is $0.46\,\mathrm{\mu eV/\sqrt{Hz}}$ at $1\,\mathrm{Hz}$ and $6.77\,\mathrm{\mu eV/\sqrt{Hz}}$ at $10\,\mathrm{mHz}$.}
\end{figure}

\begin{figure}[!t] 
\centering
\includegraphics[width=\linewidth] {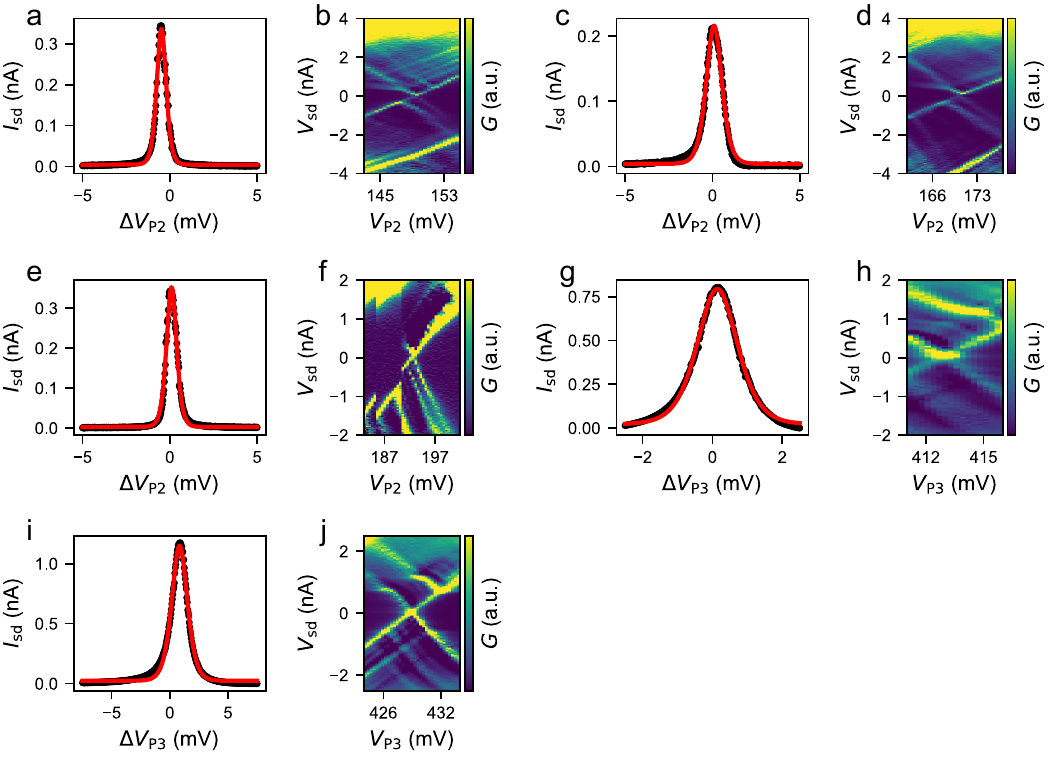}
\caption{\textbf{Key parameters for quantifying charge noise via Coulomb peak tracking measurements.} \textbf{a}, Coulomb oscillation curves of Peak 1 in device B (black points) and fitting curve using the hyperbolic secant function (red curve). The peak position is extracted from curve fitting (see Methods). \textbf{b}, Coulomb diamond of Peak 1 in device B for the extraction of the gate level arm. \textbf{c-f}, Coulomb oscillation curves and Coulomb diamonds for Peak 2 (c, d) and Peak 3 (e, f) in device B. \textbf{g-j}, Coulomb oscillation curves and Coulomb diamonds for Peak 1 (g, h) and Peak 2 (i, j) in device C.}
\end{figure}

\begin{figure}[!t] 
\centering
\includegraphics[width=\linewidth] {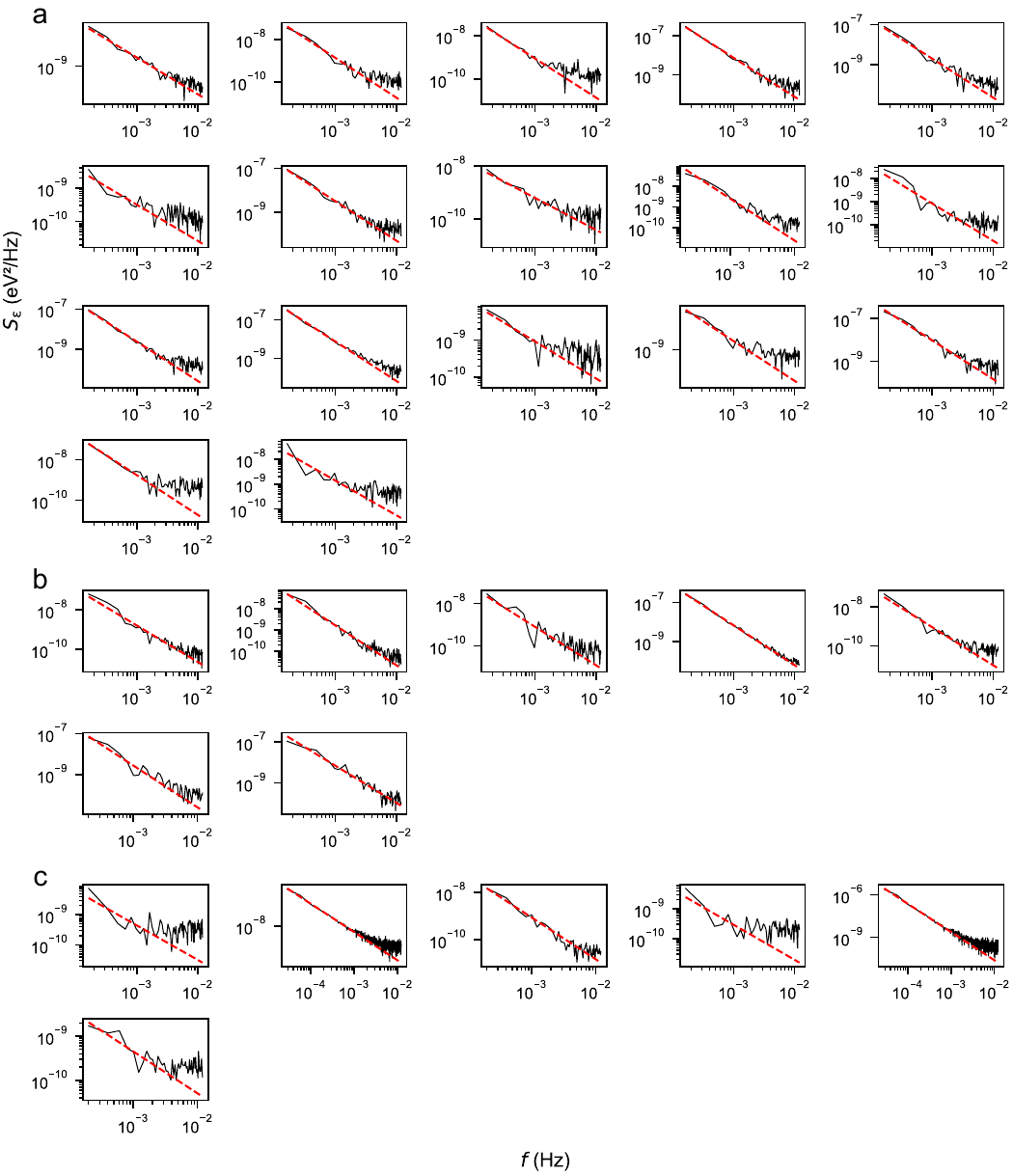}
\caption{\textbf{Charge noise spectra of three devices measured via Coulomb peak tracking.} The spectra of devices A, B and C are shown in \textbf{a}, \textbf{b} and \textbf{c}, respectively. The black traces are data and the red dashed lines are fitting curves via the formula $1/{f^\alpha}$. The extracted charge noise data from these measurements are presented in Fig. 5c. The average charge noise is $5.92\,\mathrm{\mu eV/\sqrt{Hz}}$ at $10\,\mathrm{mHz}$ and $46.48\,\mathrm{\mu eV/\sqrt{Hz}}$ at $1\,\mathrm{mHz}$.}
\end{figure}

\begin{figure}[!t] 
\centering
\includegraphics[width=0.6\linewidth] {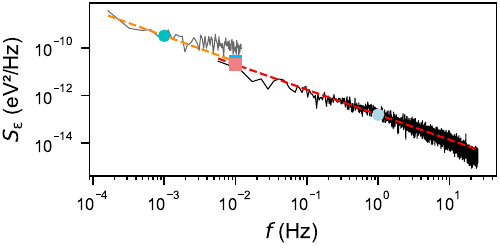}
\caption{\textbf{Comparison of charge noise data extracted with two methods (current spectroscopy versus Coulomb peak tracking).} The black curve shows representative data of $S_{\epsilon}$ as a function of $f$ obtained from current spectroscopy, and the red dashed line is corresponding fitting curve via the formula $1/{f^\alpha}$ with the frequency range of $5\,\mathrm{mHz}-10\,\mathrm{Hz}$. Light blue dot and light coral square mark the values of $S_{\epsilon}$ in the fitting curve at $1\,\mathrm{Hz}$ and $10\,\mathrm{mHz}$, respectively. The grey curve is representative data of $S_{\epsilon}$ as a function of $f$ obtained from Coulomb peak tracking, and the orange dashed line is the fitting curve via the formula $1/{f^\alpha}$ with the frequency range of $0.2-2\,\mathrm{mHz}$. Green dot and blue square mark the values of $S_{\epsilon}$ in the fitting curve at $1\,\mathrm{mHz}$ and $10\,\mathrm{mHz}$, respectively. The combine data nearly follows a simple form of $1/{f^\alpha}$, and the extracted charge noise amplitude at $10\,\mathrm{mHz}$ via the two methods is also comparable ($4.45\,\mathrm{\mu eV/\sqrt{Hz}}$ versus $5.16\,\mathrm{\mu eV/\sqrt{Hz}}$).}
\end{figure}

\clearpage